\documentclass[prc,twocolumn,aps,reprint,nofootinbib,showpacs,superscriptaddress]{revtex4}

\usepackage{graphicx}
\usepackage{amssymb,amsmath,amstext,amsthm,amsfonts}
\usepackage[bookmarks,pdfhighlight=/O,pdfstartview=FitH]{hyperref}
\usepackage{slashed}
\usepackage[usenames]{color}

\newcommand{\GeV}{\; \mathrm{GeV}}
\newcommand{\MeV}{\; \mathrm{MeV}}
\newcommand{\eV}{\; \mathrm{eV}}

\newcommand{\dd}{\mathrm{d}}

\begin{document}

\title{Neutrino energy reconstruction in quasielastic-like scattering\\ in the MiniBooNE and T2K experiments}
\author{O. Lalakulich%
\email[Contact e-mail: Olga.Lalakulich@theo.physik.uni-giessen.de]}
\author{U. Mosel}
\affiliation{Institut f\"ur Theoretische Physik, Universit\"at Giessen, Germany}
\author{K. Gallmeister}
\affiliation{Institut f\"ur Theoretische Physik, Johann Wolfgang Goethe-Universit\"at Frankfurt, Germany}

\begin{abstract}
\begin{description}
\item[Background] Neutrino oscillation probabilities, which are being measured in long-baseline experiments,
depend on neutrino energy. The energy in a neutrino beam, however, is broadly
smeared so that the neutrino energy in a particular event is not directly known, but must be reconstructed from final state properties.
\item[Purpose] To investigate the effects of different reaction mechanisms on the energy-reconstruction method widely
used in long-baseline neutrino experiments taking also pion production into account, and to clarify how the oscillation signal depends on the energy reconstruction.
\item[Methods] The Giessen Boltzmann-Uehling-Uhlenbeck (GiBUU) transport model is used for a detailed study of neutrino-nucleus events.
\item[Results] The difference between the true-QE and QE-like cross sections in the MiniBooNE experiment is investigated in detail.
It is shown that fake QE-like events lead to significant distortions in neutrino energy reconstruction.
Flux-folded  and unfolded cross sections for QE-like scattering are calculated as functions of both true and reconstructed energies.
Flux-folded momentum transfer distributions are calculated as functions of both true
and reconstructed momentum transfer. Distributions versus reconstructed values are compared with the experimental data.
Also presented are the conditional probability densities of finding a true energy for a given reconstructed energy.
We show how the energy reconstruction procedure influences the measurement of oscillation parameters in T2K experiment.
\item[Conclusions] For the reconstruction procedure based on quasielastic (QE) kinematics all other reaction channels besides true-QE
scattering show a  shift of the reconstructed energy towards lower values as compared to the true energy.
In the MiniBooNE and T2K experiments this shift is about 100 - 200 MeV and depends on energy.
The oscillation signals are similarly affected. These uncertainties may limit the extraction
of a $CP$ violating phase from an oscillation result.
\end{description}
\end{abstract}

\maketitle

\section{Introduction}

The extraction of neutrino oscillation parameters necessarily requires the knowledge of the neutrino energy. Since neutrino beams are always produced as secondary decay products their energy is not sharp, but widely distributed. The energy of the incoming neutrino in a given event thus has to be reconstructed from observed properties of the final state. Often QE scattering is used for this reconstruction, because for this process the incoming energy can be uniquely inferred from an observation of just the outgoing lepton if the target is a nucleon at rest. This becomes considerably more complicated, however, when nuclear targets are used. For these, the difficulty is twofold. First, the reaction process of QE scattering must be unequivocally identified. Second, even then nuclear effects can smear out the reconstructed energy. The first difficulty is the more serious one, since other reaction mechanisms may look indistinguishable in the experiment \cite{Delorme:1985ps,Marteau:1999jp}. Furthermore, final state interactions make it very difficult to identify the initial QE scattering on a bound, Fermi-moving nucleon inside the nucleus. This is even more so if no outgoing nucleons are observed, as is at present the case in all experiments using Cherenkov detectors.

Recently quasielastic (QE) neutrino scattering has attracted a lot of attention because
the cross sections reported lately~\cite{Gran:2006jn,Katori:2009du,AguilarArevalo:2010zc}
by the K2K and MiniBooNE  collaborations at $E_\nu<2\GeV$ are $30\%-40\%$ higher than the cross section measured
by old bubble-chamber experiments (ANL, BNL, FNAL, CERN, and IHEP) in the 70s and 80s. These higher data could be well described within a relativistic Fermi gas model only by assuming an axial mass that was significantly higher than the world average value of $M_A \approx 1.0 \GeV$ \cite{Gran:2006jn,AguilarArevalo:2007ab,AguilarArevalo:2010cx}.
The situation is complicated by another recent experiment -- NOMAD~\cite{Lyubushkin:2008pe}
-- which at $E_\nu>5\GeV$ reports  $M_A=1.05\pm 0.08 \GeV$ and
cross sections which are in agreement with the old measurements.
The Minerva experiment, operating with a higher energy neutrino flux,
also does not derive a high $M_A$ from its preliminary results \cite{McFarland:2011jd}.

At first sight, the modern experiments have the advantage of huge statistics
with millions of events recorded.
The complication arise, however,  from the fact that these experiments all use nuclei as
targets. All the measurements are thus influenced by nuclear effects.
These affect primarily the event identification and with it the energy reconstruction. 
This is crucial for the extraction of the cross section as a function of neutrino energy \cite{AguilarArevalo:2010cx} and influences the extraction of oscillation parameters \cite{Abe:2012gx}.

The true charged current (CC) QE scattering is defined as neutrino scattering on a bound neutron in the nucleus,
resulting in a muon and a proton:
\[
\mbox{true QE:} \qquad \nu n \to \mu^- p
\]
Such events can be identified quite well in a tracking detector, but are impossible to identify in a Cherenkov detector,
because it is ''blind`` to outgoing neutrons and low-energy outgoing protons.
Thus, in a Cherenkov detector (e.g.\ MiniBooNE and T2K~\cite{Abe:2012gx}), the signal is defined as
a single Cherenkov ring from the outgoing muon, which can also
be identified by its decay electron. No further rings, possibly originating from pions or other mesons,
should appear in such an event.   Thus, CC QE-like events are defined as those with
1 muon, 0 mesons and any number of nucleons in the final state.

Not all events that appear QE in the Cherenkov detector have a true QE origin\cite{Delorme:1985ps,Marteau:1999jp,Leitner:2010kp}. In the MiniBooNE and T2K energy regimes one complication arises
from processes in which a pion or a $\Delta$ resonance is produced in the initial neutrino vertex, e.g., $\nu p \to \mu^- p \pi^+$ or $\nu p \to \mu^- \Delta^{++}$,
and then is absorbed in the nucleus through final state interactions.
Thus the pion is not seen in the final state (the so-called 'stuck-pion' event) and such event is counted as QE-like even though it is of not true-QE origin.
In an earlier publication \cite{Leitner:2010kp} we have shown that there is indeed a strong entanglement of QE scattering and pion production events
and that, consequently, Cherenkov detectors always see too high a cross section for QE scattering.  A second complication is the presence of multi-nucleon events in which the incoming neutrino interacts with, e.g.,\ 2 nucleons (so-called 2p-2h events). Both of these complications lead to the so-called fake QE events. The ''measured'' QE-like cross section is then contaminated by these fake QE events. On the other hand, a nucleon produced in the initial QE vertex may rescatter in the nucleus
and produce a pion in the final state. This event will be disregarded as QE-like, even though it originates in a true-QE scattering process.

\begin{figure}[!hbt]
\includegraphics[width=\columnwidth]{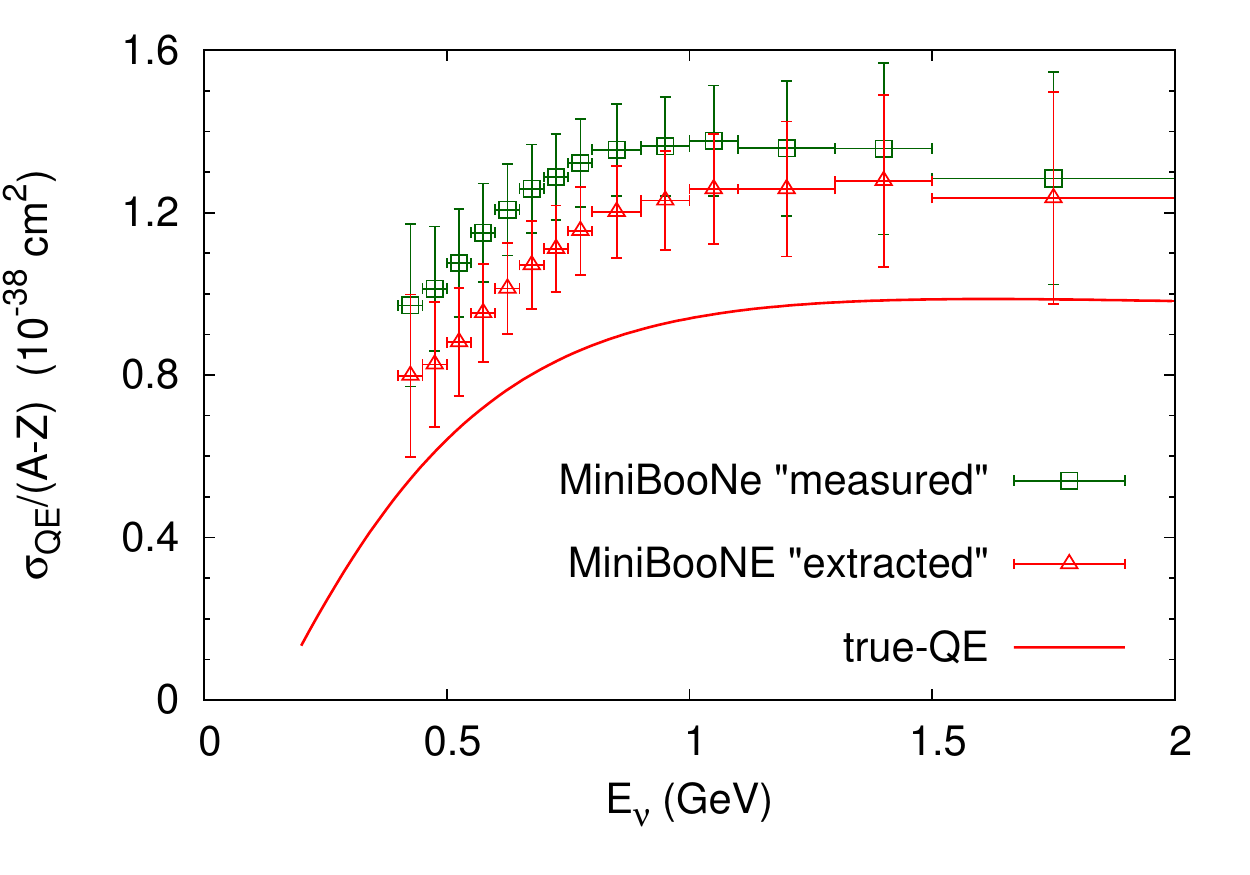}
\caption{(Color online) The squared points give the ``measured'' QE-like cross section
obtained by the MiniBooNE Collaboration~\cite{AguilarArevalo:2010zc}.
After subtracting from this the ``QE-like background'' (see text) the ``extracted'' cross section
shown by the triangular points is obtained~\cite{AguilarArevalo:2010zc}.
All data are plotted vs reconstructed energy.
The solid line gives the result for the true-QE cross section, obtained in a
GiBUU calculation using the world average value for the axial mass of $M_A = 1 \GeV$.}
\label{fig:QE-versusTrue}
\end{figure}

The experimental groups try to account for nuclear effects by using event-generators.
Thus, the final 'data' for QE scattering invariably contain some model dependence and may suffer from imperfections in the models used. For example, the widely discussed large values for the axial mass ($M_A \approx 1.3 \GeV$) obtained by the MiniBooNE collaboration are nowadays believed to be due to 2p-2h excitations (see Refs.~\cite{Martini:2011wp,Nieves:2011yp} and references therein). These excitations were not contained in the generator used and thus could not be removed from the data set. In addition, one has to be aware that in the MiniBooNE data shown in Fig. \ref{fig:QE-versusTrue} neither the QE cross section itself nor the neutrino energy have been measured directly. Instead only the flux-averaged double-differential (with respect to outgoing lepton variables) QE-like cross section can directly be measured. Even the energy-separated QE-like cross section (full squares in Fig.~\ref{fig:QE-versusTrue}, denoted as ''measured``) as a function of neutrino energy is model-dependent; the same is naturally true for the QE-extracted cross section (open triangles in Fig.~\ref{fig:QE-versusTrue}, denoted as ''extracted''). In both cases the energy-dependence has been reconstructed. The stuck-pion processes should account for most of the difference between these two data sets.

It is the purpose of this paper to explore the full QE-like cross section that includes the events from various origins and in particular also from pion production.
These are clarified in Sec.~\ref{QElike}. 
Special attention is given to the 2p-2h processes, which are discussed in Sec.~\ref{2p2h}.
In Sec.~\ref{energy-QEKinematics} we consider the sensitivity of the energy reconstruction method used by Cherenkov detectors such as MiniBooNE to the events of each  origin in a comprehensive way. In particular, we explore the dependence of the energy reconstruction not just on 2p-2h and $\Delta$ events, but also on QE-like contributions from pion background, higher resonances and Deep Inelastic Scattering (DIS).
Reconstruction of momentum transfer is discussed in Sec.~\ref{Q2-QEKinematics}.
The influence of energy reconstruction procedure on measurement of oscillation parameters in the T2K experiment is illustrated in Sec.~\ref{T2K}.
At the end we summarize our findings.


\section{QE-like cross sections \label{QElike}}

\subsection{Theoretical Method}

For our investigations of nuclear effects on the QE-like and extracted QE cross sections we use the Giessen Boltzmann-Uehling-Uhlenbeck (GiBUU) transport model. Its theoretical foundation and details of the practical implementation are described in detail in Ref.\ \cite{Buss:2011mx}. In this model we solve the approximate Kadanoff-Baym equations for the time development of the spectral phase space distributions of nucleons and mesons after a nucleus has interacted with an incoming particle. The model, which has been widely tested on a variety of nuclear reactions, is an event generator that delivers at the end, i.e., after all final state interactions, the four-momenta of all asymptotically free on-shell particles. It contains all the relevant processes: QE scattering, background pion production,  and nucleon resonance decays are included; all resonance properties are taken either from the Manley analysis \cite{Manley:1992yb} and (for electromagnetic vector couplings) from the MAID analysis \cite{Tiator:2009mt,Leitner:2008ue}. The model has recently been extended by also including initial two-particle-two-hole (2p-2h) interactions \cite{Lalakulich:2012ac}. At higher energies pQCD expressions encoded in the high-energy event generator \textsc{pythia} are used \cite{Lalakulich:2012gm}. The model thus contains all the relevant processes for the initial neutrino interaction.   All calculations reported in this paper have been performed with an axial mass $M_A = 1 \GeV$ for QE scattering; model II from Ref.\ \cite{Lalakulich:2012ac} is used for the 2p-2h contributions.
Once the particles are produced in the initial vertex they are being transported out of the nucleus. During this process elastic and inelastic scattering can take place, including charge transfer, particle production, and absorption.

\subsection{True QE and QE-like cross sections}

\begin{figure}[bht]
\includegraphics[width=\columnwidth]{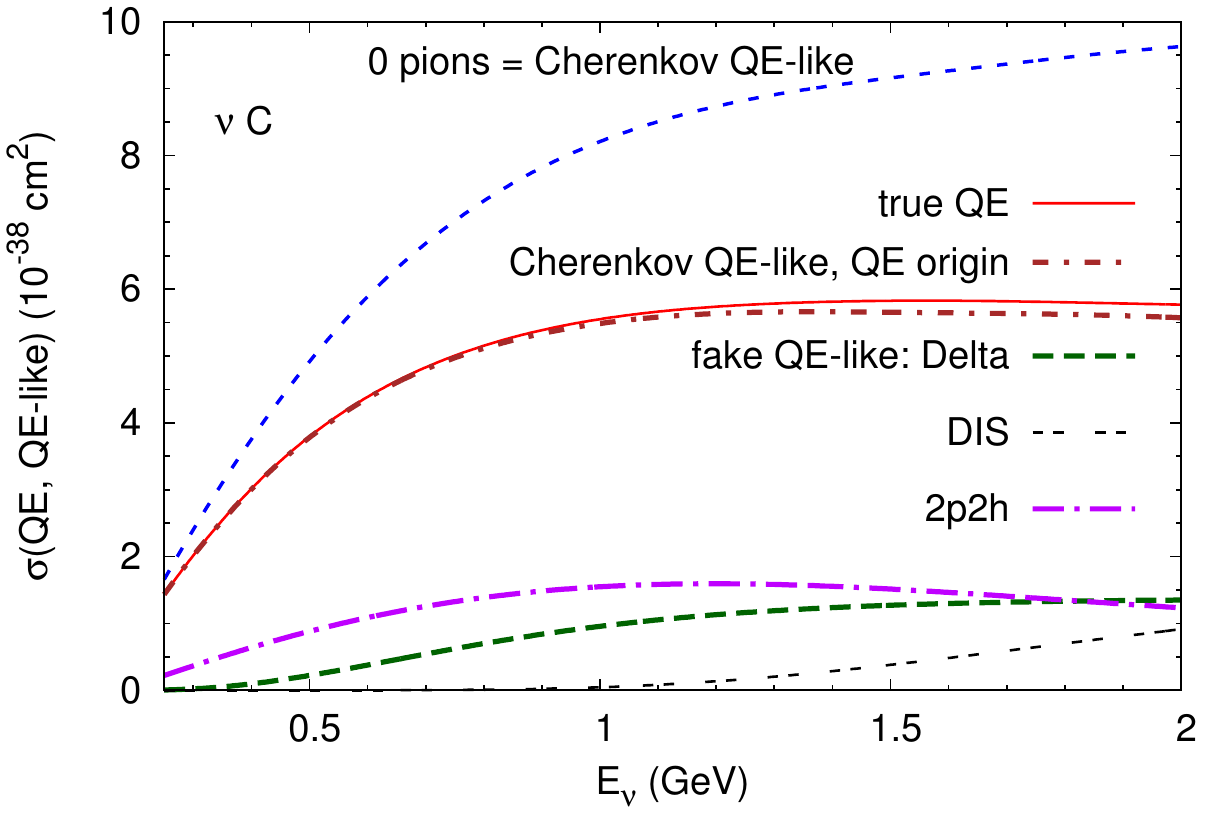}
\caption{(Color online) True QE (solid curve) and QE-like (short-dashed curve) cross sections for neutrino scattering off carbon.
Model II from Ref.~\cite{Lalakulich:2012ac} has been used for the 2p-2h interactions.}
\label{fig:carbon-QE-true-like}
\end{figure}

Figure~\ref{fig:carbon-QE-true-like} shows true-QE  and QE-like  cross sections versus neutrino energy as obtained in the GiBUU model for neutrino CC scattering off carbon. The QE-like events have been identified using the same method as that employed with Cherenkov detectors.
As one can see from Fig.~\ref{fig:carbon-QE-true-like}, the Cherenkov detector sees almost all true QE events, but also a large part
of the fake QE-events.  Such events are called ''QE-like background`` by MiniBooNE.
So the resulting QE-like cross section is about 30\%-40\% higher than  the true-QE one.
The fake events, as shown in Fig.~\ref{fig:carbon-QE-true-like}, may come from $\Delta$ and other resonances decaying to mesons, one-pion background,
and, at higher energies, from DIS  when the outgoing mesons  are absorbed in the nucleus during the Final State Interactions (FSI).  In addition, there may be events present in which the initial interaction involved more than one nucleon (the so-called 2p-2h processes; see the discussion in Sec.~\ref{2p2h}).

In MiniBooNE, the QE-like background is not directly measured, but obtained  from the NUANCE neutrino event generator which in turn was adjusted to the measured pion yields \cite{AguilarArevalo:2010cx}; this generator contained no 2p-2h processes. This background  should  correspond to the sum of fake events
originating from resonances, one-pion background, and DIS in the GiBUU model.
Subtracting the QE-like background from the ''measured`` QE-like cross section,
MiniBooNE presented its QE result (which is meant to be sensitive to the axial mass) as ''extracted`` cross section (cf. Fig.~\ref{fig:QE-versusTrue}).

The neutrino energy for each event has been reconstructed from muon observables only assuming quasifree QE scattering on a nucleon at rest.
This reconstruction is necessary because in any experiment the neutrino  beam involves a broad energy distribution, and the true energy for a given event is thus not known. It can lead to inaccuracies in the reconstructed energies that are larger than previously assumed if the actual reaction process is not correctly identified.

We note that for inclusive cross sections (especially at high neutrino energy) an identification of a particular reaction mechanism is not necessary so that the experiments can rely on calorimetric measurements of the energy of final hadrons (MINOS, NOMAD). However, even in this case large corrections of the actually measured energies are needed because due to rescattering a large part of the initial signal can go unobserved \cite{Kordosky:2006gt,Kordosky:2007tu}


\subsection{Influence of  2p-2h interactions \label{2p2h}}

As we see from Fig.~\ref{fig:QE-versusTrue}, removing all the stuck-pion QE-like events does not lead to an agreement with the conventional theoretical calculation of true-QE cross section.
The extracted cross section still is considerably higher than calculations using the world average axial mass of 1 GeV.
The remaining difference can be attributed to  a significant amount of non-QE many-body excitations in the QE-like cross section; for a compact
review of other possible explanation see the introduction in \cite{Lalakulich:2012ac}.
Since the NUANCE generator used by MiniBooNE does not contain 2p-2h excitations, they are not subtracted as a part of the QE-like background, and thus are still present in the measured cross section.

By combining the random phase approximation (RPA)  with a calculation of 2p-2h contributions Martini et al.\ have obtained a good description of the MiniBooNE data \cite{Martini:2009uj,Martini:2010ex,Martini:2011wp}. As expected, the RPA correlations have the most effect at forward angles where the squared four-momentum transfer $Q^2$ to the nucleus is small. They die out with increasing angle and with decreasing muon energy, i.e.\ increasing energy transfer. These results have been confirmed by detailed calculations of the Valencia group \cite{Nieves:2011pp,Nieves:2011yp} and the effects of such interactions have been explored in detail in Ref.~\cite{Lalakulich:2012ac}(see also Ref.~\cite{Sobczyk:2012ms}).

Since all measurements involve a flux-average over a usually broad band of incoming neutrino energies we have developed a short-cut to the full theoretical treatment of such 2p-2h processes by parametrizing a flux-averaged matrix element. This method has allowed us to implement these processes in an actual event generator, GiBUU, and to investigate observable consequences of these initial 2p-2h interactions mainly on knock-out nucleons. In Ref.~\cite{Lalakulich:2012ac} we have shown that the measured double-differential cross sections (corrected for stuck-pion events) could be described quite well so that our method seems to contain the relevant physics. There we had used a model for the 2p-2h part of the hadron tensor that consisted of the transverse projector modified by an explicit energy-dependence  (model II in Ref.~\cite{Lalakulich:2012ac}). We model in this way the fact that two-body terms contribute to the transverse strength over the quasielastic region, and become sizable for energy transfers beyond the QE peak \cite{Shen:2012xz}. In Ref.~\cite{Amaro:2010iu} it has been shown within the relativistic Fermi gas model that in electron scattering 2p-2h correlations can also lead to longitudinal strength some of which is due to non-meson-exchange-current (MEC) correlations.  These correlations also contribute to the spectral function of nucleons and it is thus not clear how much of them are already included when working with bound nucleons, as we do in the GiBUU calculation. In the absence of a consistent theoretical study of this problem for neutrino reactions, therefore, our ansatz for the hadron tensor being purely transverse cannot be more than an educated guess. As shown in ref.~\cite{Lalakulich:2012ac} the main effect of this predominance of transversality in the 2p-2h interaction is an increased strength at backward muon angles. There is, however, only a small effect of this increase on the reconstructed energies.

\begin{figure}[bht]
\includegraphics[width=\columnwidth]{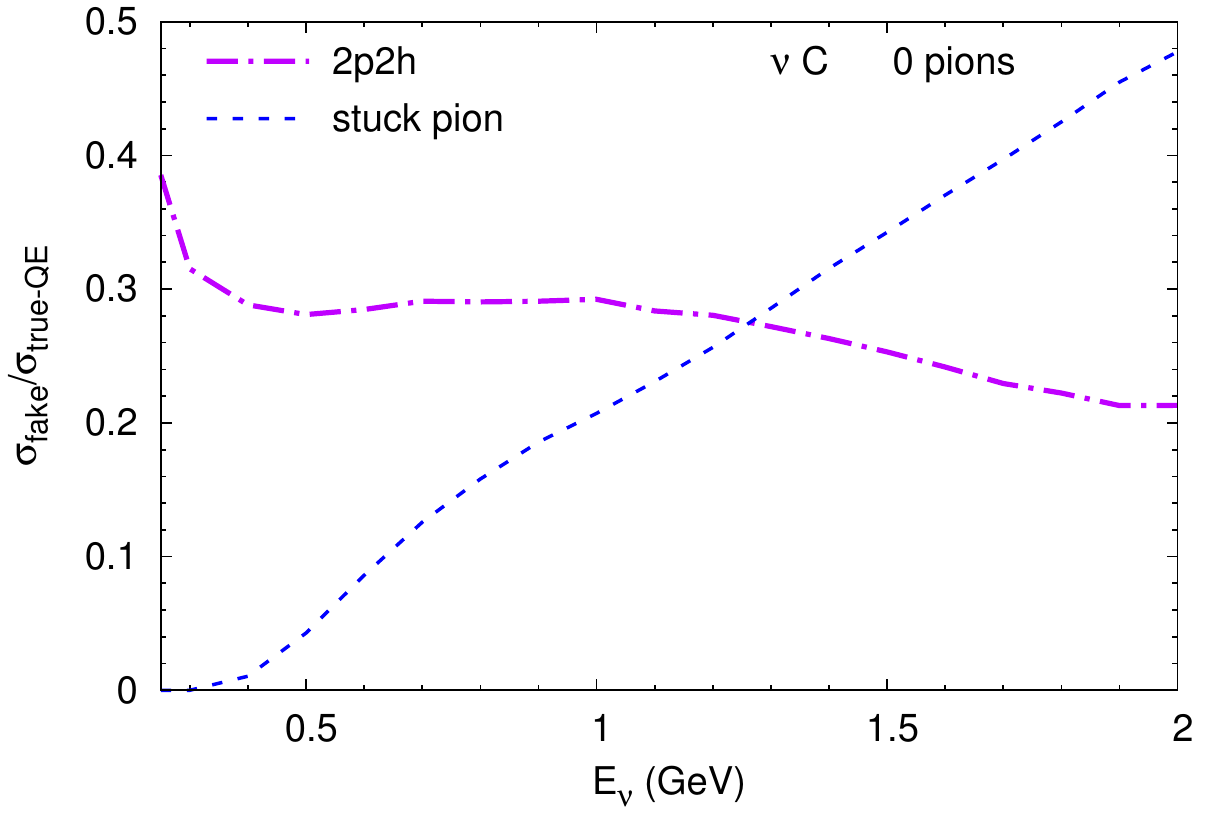}
\caption{(Color online) Ratios of fake QE-like cross sections of 2p-2h and stuck-pion origins to the true-QE one. }
\label{fig:carbon-ratios-to-QE}
\end{figure}

The 2p-2h cross section leading to fake QE-like events is also shown in Fig.\ \ref{fig:carbon-QE-true-like}. It is seen that at low energies, $E_\nu<0.4 \GeV$, fake events come only from 2p-2h contributions because at these low energies processes involving resonance excitations or pions are kinematically suppressed. Above this energy the main contribution comes
from the 2p-2h mechanism and $\Delta$ excitation followed by its decay. With increasing energy the contribution of  fake events grows, because new channels
(production of higher baryonic resonances, DIS) open up.
The relative contributions of fake events of 2p-2h (dash-dotted line) and stuck-pion (dashed line) origins are shown in Fig.~\ref{fig:carbon-ratios-to-QE}.
While the 2p-2h contribution is quite flat at about 20-30\%, the pion contribution rises with energy, reflecting the threshold for pion production.

\begin{figure}[!hbt]
\includegraphics[width=\columnwidth]{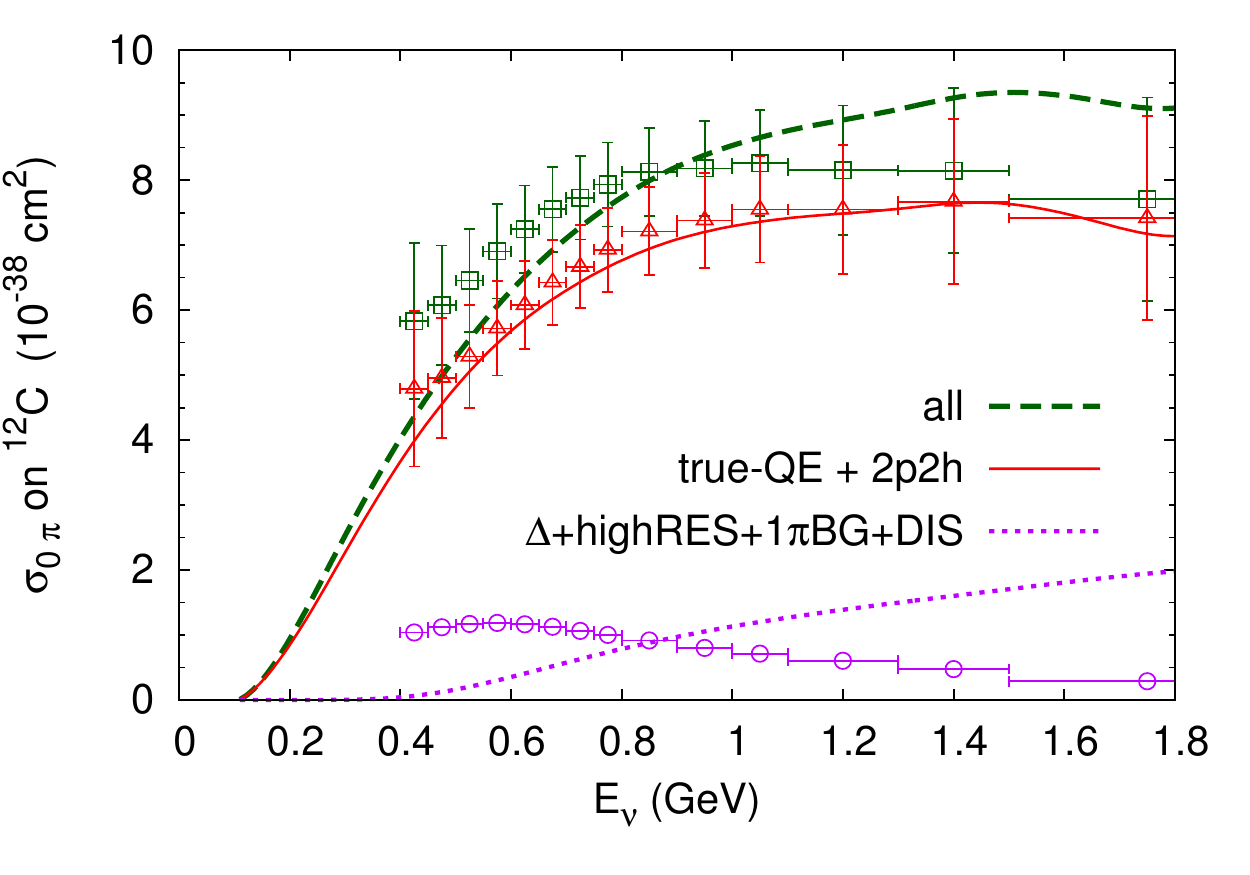}
\caption{(Color online)
QE-like cross section originating from QE and 2p-2h processes only (solid line) and from all processes (dashed line) within
the GiBUU calculations. Measured and extracted MiniBooNE data points are the same as in Fig.~\ref{fig:QE-versusTrue}.
The difference between them (open circles) is compared with the GiBUU stuck-pion cross section (dotted line).
All data are plotted vs reconstructed energy, whereas the theoretical curves are plotted vs true neutrino energy. }
\label{fig:all-versusTrue}
\end{figure}

In Ref.~\cite{Lalakulich:2012ac}  we had determined the matrix element such that the sum of true-QE and 2p-2h contributions
fitted the extracted MiniBooNE data. This is shown in Fig.~\ref{fig:all-versusTrue}, where the solid
(``true-QE + 2p2h'') line is the GiBUU model calculation that includes only true-QE and 2p-2h cross sections.
Even with this fit, as illustrated in Fig.~\ref{fig:all-versusTrue}, the measured data points  still do not agree with
our curve for the total QE-like cross section.
The latter  is shown by the dashed (``all'') line and  includes all processes that lead to a QE-like final state.

As shown in the previous section, the absolute contribution of fake stuck-pion QE-like events (that is, the difference between the dashed and the solid curves in Fig.~\ref{fig:all-versusTrue}, also shown as dotted curve)  is zero for $E_\nu<0.4 \GeV$ and slowly grows with increasing energy.
The MiniBooNE results (open circles), however, show quite a different picture.
The contribution of fake events is  largest at low energies and decreases further as energy grows (open circles in Fig.~\ref{fig:all-versusTrue}). The theoretical ``all'' and ``true-QE + 2p2h'' curves do not agree with the data; both have a noticeably different shape. As we will show later in this paper, the resolution of this seeming contradiction lies in the fact that in Fig.~\ref{fig:all-versusTrue}  the data are plotted versus reconstructed energy whereas the calculated curves are all plotted versus true energy.

\section{Energy reconstruction based on QE kinematics}
\label{energy-QEKinematics}
To resolve the contradiction shown  in Fig.~\ref{fig:all-versusTrue}, let us  consider
the energy reconstruction procedure used by MiniBooNE and its influence not just on the QE scattering, but also on the QE-like cross sections.
As was shown already in \cite{Martini:2012fa,Nieves:2012yz,Lalakulich:2012ac}, a 2p-2h interaction, when leading to a final state with zero pions and
thus recorded as QE-like event, is on average recorded with a reconstructed energy lower than the true energy.

For QE scattering on a nucleon at rest the incoming neutrino energy is directly linked to the kinematics of the outgoing lepton and is thus
known when lepton angle $\theta_\mu$ and energy $E_\mu$ are measured. Therefore, the formula used by MiniBooNE for the energy reconstruction
is based on the assumption of QE scattering on a nucleon at rest \cite{AguilarArevalo:2008yp} even though nuclear targets with binding and Fermi motion are used.
The reconstructed (rec) neutrino energy is defined as
\begin{equation}   \label{Ereconstr}
E_\nu^\mathrm{rec} = \frac{2(M_n - E_B)E_\mu - (E_B^2 - 2M_n E_B + m_\mu^2 + \Delta M^2)}{2\left[M_n - E_B - E_\mu + |\vec{k}_\mu| \cos \theta_\mu\right]} ~.
\end{equation}
Here $M_n$ is the mass of the neutron, $\Delta M^2 = M_n^2 - M_p^2$, and $|\vec{k}_\mu|=\sqrt{E_\mu^2-m_\mu^2}$ is the absolute value  of the  three-momentum of the outgoing muon.
This formula, therefore, neglects any Fermi-motion effects; binding is taken into account only by a constant removal energy $E_B > 0$. It is essential to realize that
use of this formula is justified only if the reaction mechanism has been identified as being true QE scattering; admixture of any other reaction modes leads to an incorrect reconstruction of energy. In the following we will explore how large these errors actually are.

\begin{figure}[hbt]
\includegraphics[width=\columnwidth]{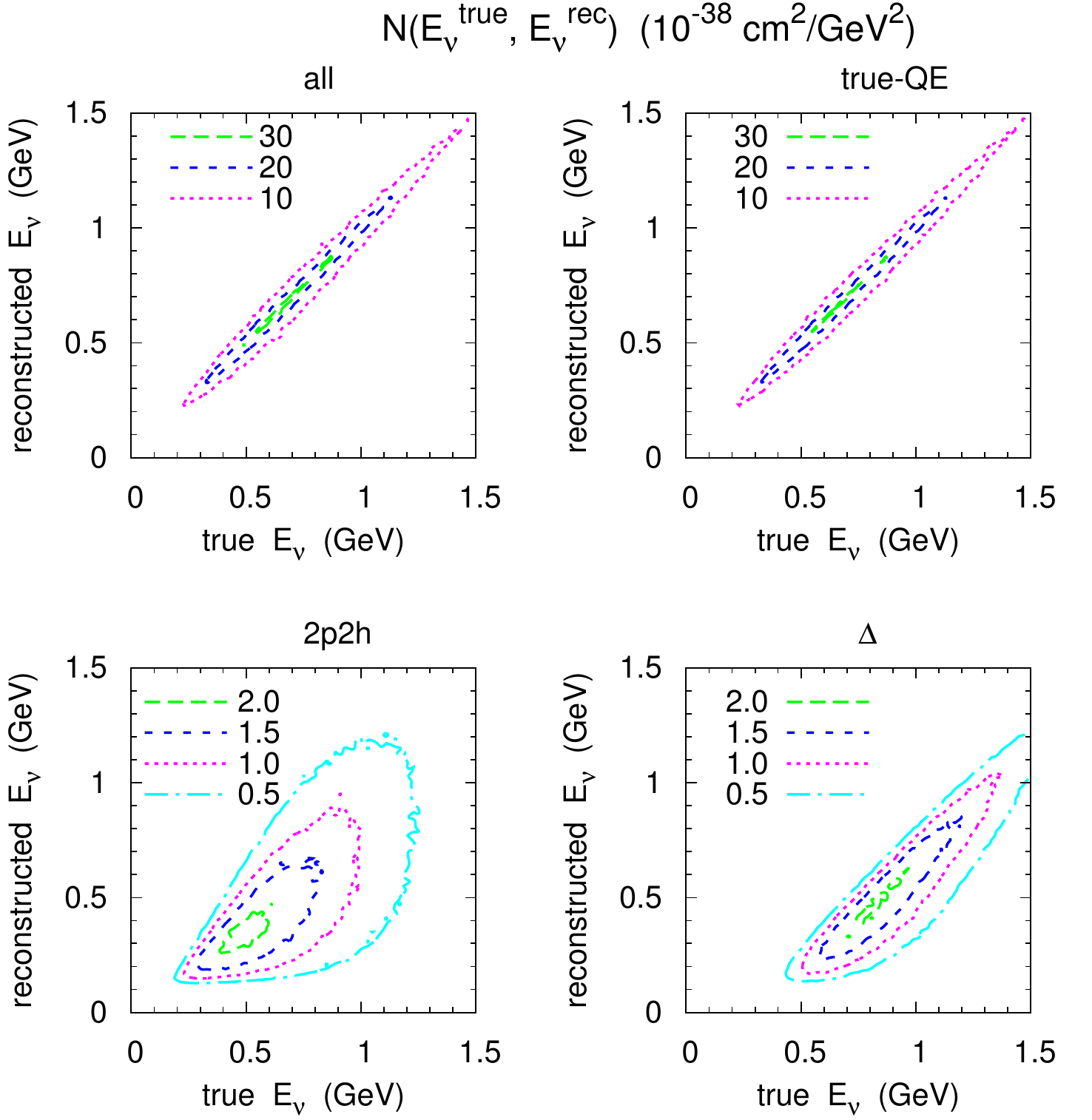}
\caption{(Color online) 2D density of the QE-like cross section $N(E^{\rm true},E^{\rm rec})$
vs true and reconstructed neutrino energies for all events and for events of various origins, all for MiniBooNE flux.}
\label{fig:events-reconstruction3D}
\end{figure}

In a numerical simulation, e.g.\, with the MiniBooNE flux, the distribution of true neutrino energies is known. In GiBUU a flux-weighted Monte Carlo sampling of true energies is then used to generate for each of them one event; from this event a reconstructed energy is calculated according to Eq.\ (\ref{Ereconstr}). This method thus directly corresponds to the method used in the experiment.
Using a sufficiently large number of true energies eventually the whole ($E^{\rm true},E^{\rm rec}$) plane is filled. The distribution of points in this plane is denoted by $N(E^{\rm true},E^{\rm rec})$, where $N$ is a two-dimensional (2D) density of the cross section.\footnote{In Ref.~\cite{Nieves:2012yz} the same value is denoted as $\frac{d\sigma}{dE_\mathrm{rec}}(E; E_\mathrm{rec}) \Phi(E)$; see Eq.(7) there.}
Integrating $N$ over $E^{\rm true}$ and $E^{\rm rec}$ gives
\begin{eqnarray}
\lefteqn{\int N(E^{\rm true},E^{\rm rec}) \, \dd E^{\rm rec} \dd E^{\rm true}} \nonumber \\
&=&  \int \phi(E^{\rm true}) \sigma_{0\pi}(E^{ \rm true})\, \dd E^{\rm true} = \langle \sigma_{0\pi} \rangle~,
\end{eqnarray}
which is just the flux averaged cross section. Here $\sigma_{0\pi}$ is the total cross section for events with zero pions in the final state, taken as a function of the true energy.

The two-dimensional density $N(E^{\rm true},E^{\rm rec})$
for the MiniBooNE flux is shown in Fig.~\ref{fig:events-reconstruction3D}.

For true-QE events the distribution is nearly symmetric about the line $E_\nu^\mathrm{true}=E_\nu^\mathrm{rec}$; the broadening comes from the Fermi motion of the bound nucleons.
For 2p-2h events at low $E_\nu^\mathrm{true}$ the reconstructed energies are mostly below the true ones; they span the full range from nearly zero to $E_\nu^\mathrm{true}$. For higher $E_\nu^\mathrm{true}$ more and more events are recorded with a reconstructed energy above the true one.
For $\Delta$ excitations, nearly all events would be recorded with reconstructed energies lower than the true ones; here the distribution is narrower than for the 2p-2h excitations due to the finite width of the $\Delta$ resonance. The same effect is even more pronounced
for excitations of higher resonances, one-pion background, and DIS events (not shown) where the absolute cross sections, however,  are lower.
Taken altogether, for all events one observes a broadening with respect to the line $E_\nu^\mathrm{true}=E_\nu^\mathrm{rec}$;
the tail towards lower reconstructed energies is most noticeable at $E_\nu^\mathrm{true} \approx 0.6-1\GeV$.
This is just the energy region where the MiniBooNE flux peaks; the situation will be similar for the T2K flux.

\begin{figure}[!hbt]
\includegraphics[width=\columnwidth]{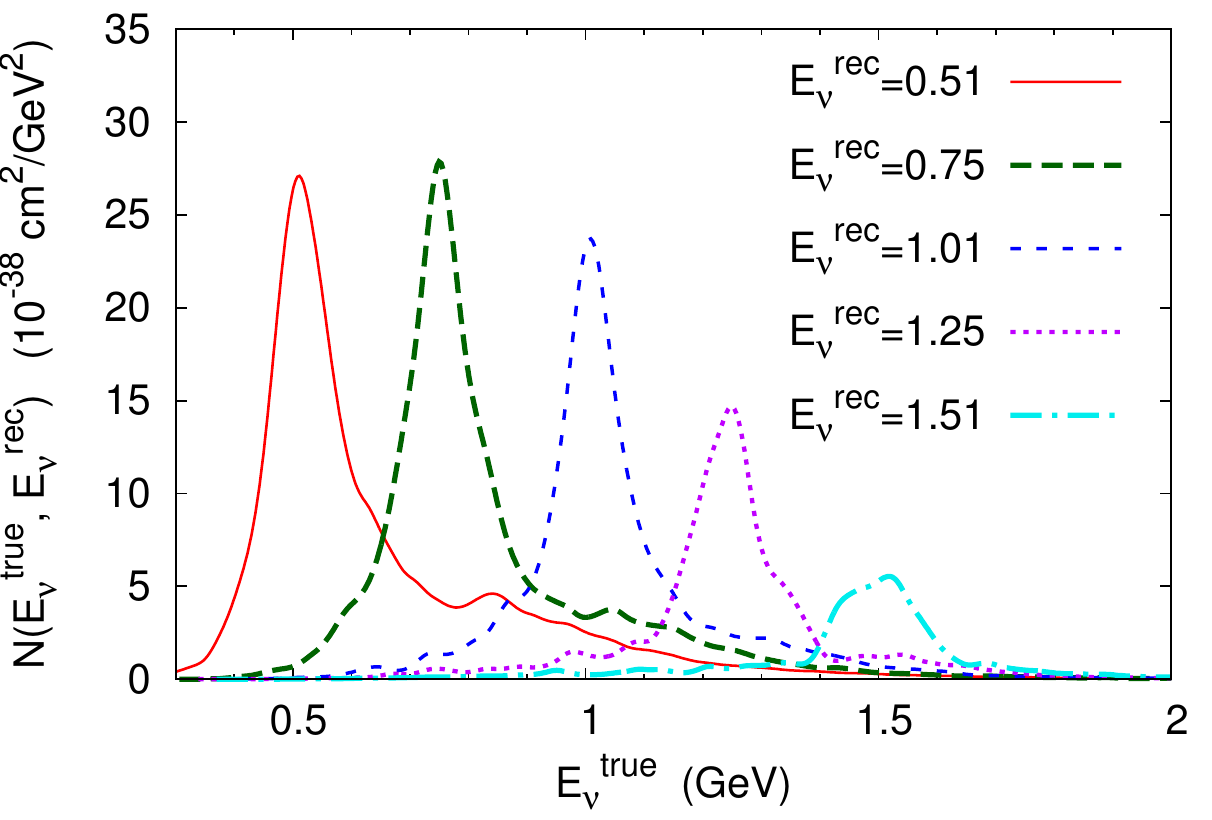}
\caption{(Color online) 2-D density of the MiniBooNE flux-averaged QE-like cross section $N(E^{\rm true},E^{\rm rec})$
as function of true energy for fixed reconstructed energies of $0.51\pm 0.01$, $0.75\pm 0.01$, $1.01\pm 0.01$, $1.25\pm 0.01$,
and $1.51\pm 0.01 \GeV$.}
\label{fig:fixedEnuRec}
\end{figure}

The distributions for given bins of reconstructed energy, i.e., horizontal cuts in Fig.\ \ref{fig:events-reconstruction3D}, are shown in Fig.~\ref{fig:fixedEnuRec} as functions of true energy.
These distributions are essential for understanding of experimental data. The curves are directly comparable with Fig.~8 in Ref.~\cite{Martini:2012fa} and  Fig.~4 in  \cite{Nieves:2012yz}. The difference is that our curves contain also the effects of pion degrees of freedom and their FSI. The main contribution to the observed distribution is in all cases given by the true-QE events, which contribute a prominent peak around the true energy; its broadening is caused by Fermi motion. In addition, the distributions have long tails towards larger true energies so that the total strength at a given true energies has sizable contributions not only from the (same) reconstructed energy, but also from lower ones.

This result is in agreement with the recent analyses by Martini et al.~\cite{Martini:2012fa} and Nieves et al.~\cite{Nieves:2012yz} for QE events. Thus, both the 2p-2h effects as well as the stuck-pion events lead to a shift of the reconstructed energy towards smaller values, or, vice versa, for a given reconstructed energy the true energy always lies higher than the reconstructed one. The effect is most pronounced at lower true energies and becomes smaller at higher energies.

Experimentally, the cross section for zero pion events, i.e., the cross section for events identified by MiniBooNE and T2K as QE-like scattering, is obtained by dividing the measured event distribution (flux-weighted cross section) at a given reconstructed energy by the flux at that same energy; we denote it by $\tilde\sigma_{\rm 0\pi}(E^{\rm rec})$. Its functional dependence on the reconstructed energy may be different from that of $\sigma_{\rm 0\pi}(E^{\rm true})$ on the true energy. On the other hand, the flux is given as a function of $E^{\rm true}$; in the procedure to determine the cross section it is read off at $E^{\rm rec}$.

Each reconstructed energy contains admixtures of many different true energies (see Fig.\ \ref{fig:events-reconstruction3D}) so that one has for the reconstructed event distribution
\begin{eqnarray}  \label{event-distrib-rec}
\lefteqn{\phi(E_{\nu}^\mathrm{rec})\tilde{\sigma}_\mathrm{0\pi}(E_{\nu}^\mathrm{rec})} \nonumber \\
&=& \int N(E^{\rm rec},E^{\rm true})\dd E_{\nu}^{\rm true} \nonumber \\
 &=& \mbox{} \int {\cal{P}}(E_{\nu}^\mathrm{rec}|E_{\nu}^\mathrm{true}) \, \phi(E_{\nu}^\mathrm{true})
 \, \sigma_\mathrm{0\pi}(E_{\nu}^\mathrm{true}) \, \dd E_\nu^{\rm true} ~,
 \end{eqnarray}
where  ${\cal{P}}(E_{\nu}^\mathrm{rec}|E_{\nu}^\mathrm{true})$ is the conditional probability density of finding a reconstructed energy $E_{\nu}^\mathrm{rec}$ for a given true energy $E_{\nu}^\mathrm{true}$. From Eq.~(\ref{event-distrib-rec}) one can read off the  probability  density ${\cal{P}}(E_{\nu}^\mathrm{rec}|E_{\nu}^\mathrm{true})$:
\begin{eqnarray}
 \displaystyle
 {\cal{P}}(\mbox{rec}|\mbox{true}) &=& {\cal{P}}(E_{\nu}^\mathrm{rec}|E_{\nu}^\mathrm{true})
\\[2mm]
\displaystyle
  &=& \frac1{\phi(E_{\nu}^\mathrm{true})  \sigma_\mathrm{0\pi}(E_{\nu}^\mathrm{true})}\,
  N(E_{\nu}^\mathrm{rec},E_{\nu}^\mathrm{true}) \ . \nonumber
\label{Prob-rec-true}
\end{eqnarray}
The probability density ${\cal{P}}(\mbox{rec}|\mbox{true})$ is a function of the reconstructed energy, while the true energy is a parameter.
It is normalized as
\[
 \int {\cal{P}}(\mbox{rec}|\mbox{true}) \dd E_{\nu}^\mathrm{rec} = 1~.
\]

Since the neutrino flux as a function of the true energy is fixed (i.e., an input in the GiBUU model, as provided by the corresponding experiments),  ${\cal{P}}(\mbox{rec}|\mbox{true})$ is independent of the neutrino flux and thus  is one and the same for all experiments using identical techniques for the event identification.
Figure~11 in Ref.~\cite{Lalakulich:2012ac}, which we will not repeat here, represents the product
${\cal{P}}(\mbox{rec}|\mbox{true}) \, \sigma_\mathrm{0\pi}(E_{\nu}^\mathrm{true})$.

The conditional probability density needed by experiment to convert the extracted event distribution, which depends on the reconstructed energy, into the true distribution, which  depends on the true energy, is discussed in detail in the Appendix.

\begin{figure}[!hbt]
\includegraphics[width=\columnwidth]{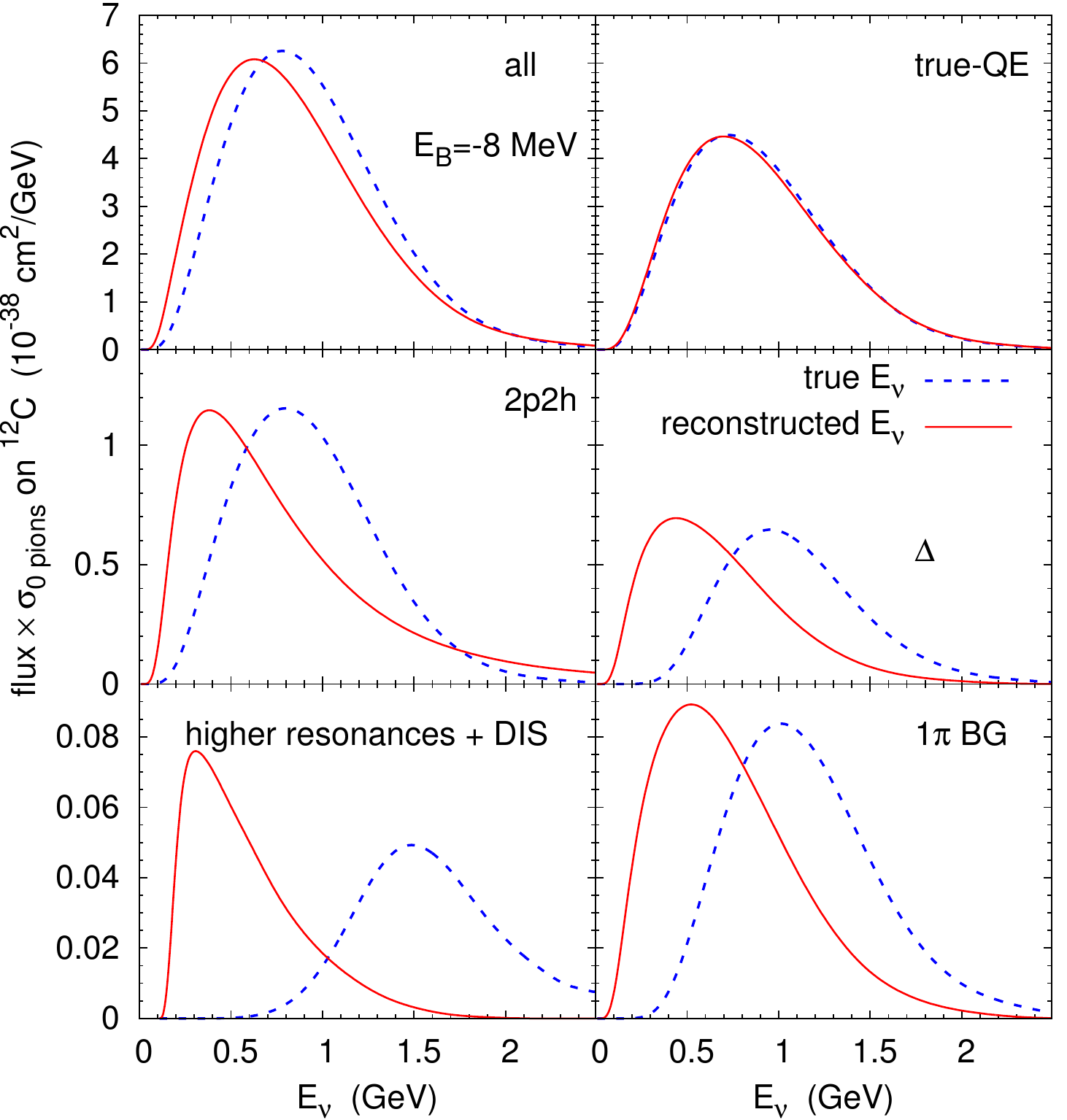}
\caption{(Color online) Event distribution of zero pion events in the MiniBooNE experiment. Shown by dashed curves are the distributions 
$\phi(E^{\rm true}) \sigma_{0 \pi}(E^{\rm true})$ of various reaction mechanisms contributing to zero pion events vs true energy 
and by solid curves the distribution $\phi(E^{\rm rec}) \tilde\sigma_{0 \pi}(E^{\rm rec})$ vs\ reconstructed energies.}
\label{fig:events-reconstruction}
\end{figure}

Figure~\ref{fig:events-reconstruction} highlights the influence of the energy reconstruction procedure on the QE-like events of various origins.
It shows the QE-like event distributions (flux-folded cross sections) versus true (solid curves) and reconstructed (dash-dotted curves) energies.
For true QE events both distributions nearly coincide. The broadening due to Fermi-motion, which is typical for fixed values of the true energy (see Fig. 11
in Ref.~\cite{Lalakulich:2012ac} as well as our discussion later in this section), has thus no visible effect.
This in turn means that, even in the presence of Fermi motion, for true-QE events the reconstruction formula
(\ref{Ereconstr}) works well as long as one aims at the cross-section measurements. This is, however, not so for
the measurements of the oscillation parameters, where the neutrino energy has to be known on an event-by-event basis
and Fermi broadening thus plays a role.

For fake QE-like events, the distributions versus true and reconstructed energies look
very different. For all origins shown in Fig.~\ref{fig:events-reconstruction} (2p-2h, $\Delta$, higher resonances, one-pion background)
the distributions versus reconstructed energy are shifted significantly to lower energies.
The largest effect comes from 2p-2h and $\Delta$ contributions, where the cross sections from fake QE-like events amount, respectively,
to 25\% and 15\% of the true-QE cross section. The overall effect for all events is that the distribution versus
reconstructed energy is shifted by about $200 \MeV$ towards lower energies with respect to the true energy.

The QE-like cross section versus reconstructed energy, $\tilde\sigma_\mathrm{0\pi}(E_{\nu}^\mathrm{rec})$, as discussed earlier,
is obtained  by dividing the measured distribution shown in Fig.~\ref{fig:events-reconstruction}  by the neutrino  flux.
The flux has a peak at $0.6\GeV$  and falls down at higher energies; so for the strongly shifted event distributions
(all fake QE-like) the effect of the reconstruction is expected to be large.

\begin{figure}[!hbt]
\includegraphics[width=\columnwidth]{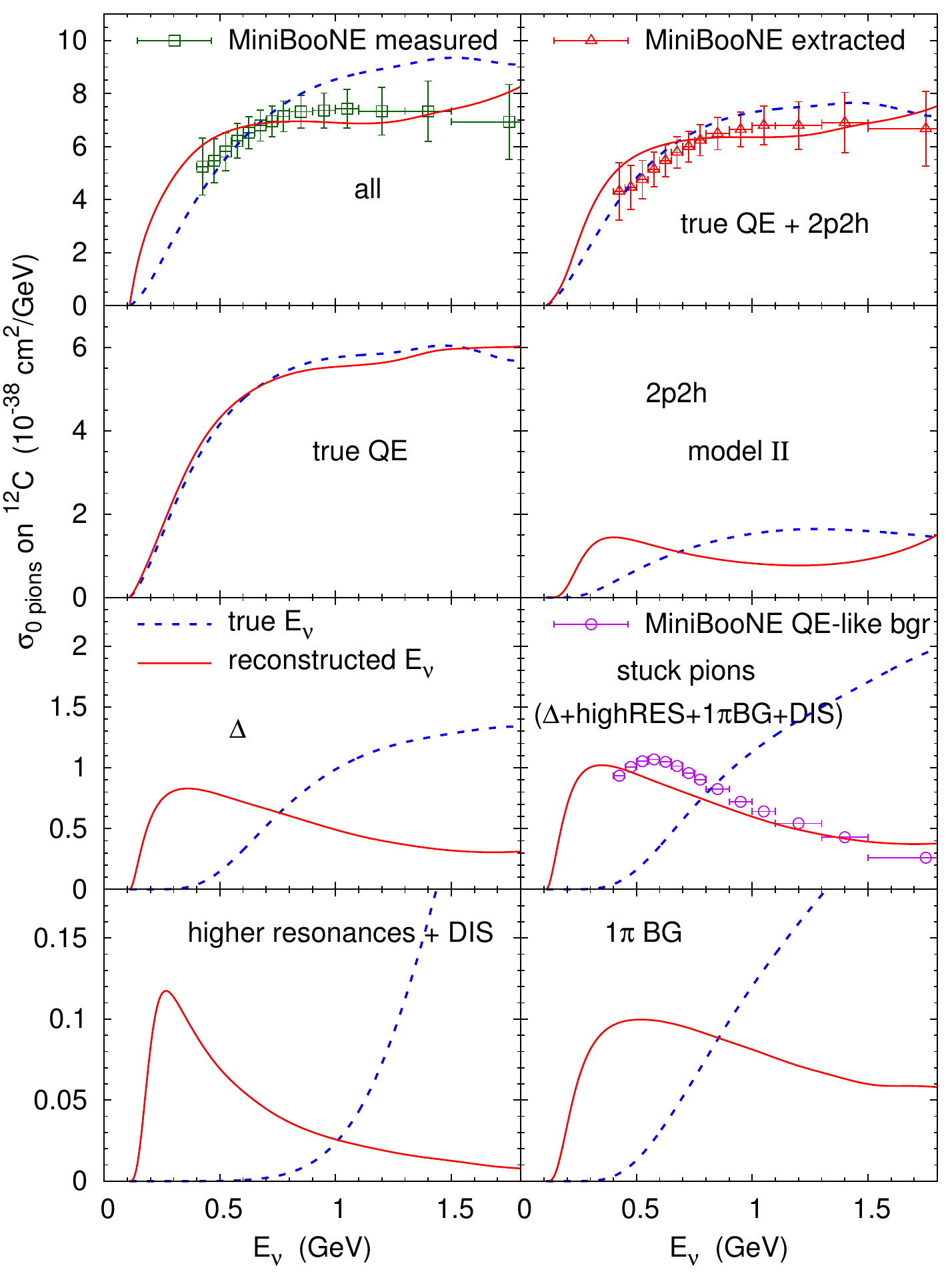}
\caption{(Color online) Various contributions to the QE-like cross sections: $\sigma_{0 \pi}$ vs true (dashed) and $\tilde\sigma_{0 \pi}$ vs reconstructed (solid) neutrino energies in the MiniBooNE experiment.  The data are multiplied by a factor 0.9. }
\label{fig:sigma-reconstruction}
\end{figure}

This expectation is indeed borne out. In Fig.~\ref{fig:sigma-reconstruction} we show the cross sections versus true and reconstructed energy separately for QE-like events of each origin.\footnote{Following Nieves et al.\ \cite{Nieves:2011yp} we have exploited the flux uncertainties in the MiniBooNE experiment and have renormalized the data in Figs.\ \ref{fig:sigma-reconstruction} and \ref{fig:Q2-events-reconstruction} by 10\% downwards.} It is seen that the energy-reconstruction method works well for the true-QE contribution, as it should. For all the other contributions, however, major differences arise and in all cases the reconstructed cross section is shifted to significantly lower energies.

In one of the panels (right, second from bottom) we also show  the GiBUU cross section for all stuck-pion processes (Delta + higher resonances + one-pion background + DIS). It should be compared with the ''QE-like background'' (open points)  as obtained by MiniBooNE using the tuned NUANCE neutrino event generator. The agreement is not perfect, but the shapes are very similar and absolute values are approximately the same. While the curve obtained by GiBUU peaks at about 300 MeV, which just corresponds to the $\Delta$ excitation energy, the MiniBooNE-NUANCE curve peaks at about 500 MeV. The difference must be due to differences between the generators used in the data extraction (NUANCE) and the GiBUU model. At first sight it is astounding that the magnitudes of the stuck-pion events as a function of true energy are roughly the same whereas it is known that the GiBUU model produces markedly less pions than measured by MiniBooNE \cite{Lalakulich:2011ne}. However, it must be remembered that the observed signal depends not only on the production, but also on the reabsorption of pions. If the tuned NUANCE produces more, but absorbs less, the result obtained here becomes understandable.

Altogether, stuck-pion events make a contribution to the fake QE-like cross section which is nearly as large as that due to 2p-2h excitations.
When these events are subtracted from the measured ones, the energy shift due to the stuck-pion events is also approximately removed by the generator.
Thus, only the energy shift induced by 2p-2h processes is present in the extracted MiniBooNE data. This justifies the
analyses based on microscopic models of 2p-2h interactions (Refs.~\cite{Martini:2012fa,Nieves:2012yz}) alone (i.e., lacking the pion contributions),
since these results are compared with the extracted QE data only.

 This now explains the puzzle, raised at the start of this paper, of why the difference between the two data sets is so large already at low energies and then decreases towards higher ones: this behavior is an artifact of the energy reconstruction. Indeed, the reconstructed-energy curve, which shows the same cross section contributions versus true energy, exhibits the expected behavior: it starts to become nonzero only for energies above about 0.5 GeV and then increases steadily with energy.

Note that for a comparison with the data in the top row in Fig.\ \ref{fig:sigma-reconstruction} only the curve for the reconstructed energy matters. We see that this curve agrees quite well with the data (reduced by 10\%) for energies above about 0.8 GeV, but overshoots them for the few experimental points at lower energies.
We take this disagreement as evidence of the presence of RPA correlations. This is in line with results of Martini et al.\ \cite{Martini:2010ex,Martini:2011wp} and Nieves et al.\ \cite{Nieves:2011pp,Nieves:2011yp} who both find that RPA correlations lower the cross sections at these lower energies, but become negligible for energies above between $0.7-1 \GeV$. An inclusion of these correlations, which are not contained in our calculations, would bring our results into very good agreement with experiment.

Summarizing the results of this section, we find that all processes other than true-QE scattering lead to a shift of the reconstructed energy, based on the assumption of true-QE scattering, towards lower energies. For both 2p-2h and pion related events this shift can be quite large (several 100 MeV). In addition the energy dependence of the cross section for various reaction mechanisms plotted versus reconstructed energy is quite different from that plotted versus true energy. While the presence of 2p-2h processes explains the abnormally large values of the axial mass obtained by the MiniBooNE Collaboration, it is interesting to look for effects of deficiencies in the energy reconstruction on neutrino oscillation properties. This will be done in Sec.~\ref{T2K}. We note that studies along these lines have already been undertaken in Refs.\ \cite{Leitner:2010kp,Martini:2012fa,Meloni:2012fq}.

\section{Momentum transfer reconstruction}
\label{Q2-QEKinematics}

The reconstruction procedure, based on true-QE kinematics and being applied to Cherenkov QE-like events, leads to  distortions not only
in the neutrino energy reconstruction,  but also in the $Q^2$ reconstruction.

Indeed, the reconstructed $Q^2$ is defined as
\begin{equation}  \label{Q2reconstr}
Q^2_\mathrm{rec} = -m_\mu^2 + 2 E_\nu^\mathrm{rec} (E_\mu-|\vec{k}_\mu| \cos \theta_\mu) \ ,
\end{equation}
using the reconstructed energy $E_\nu^\mathrm{rec}$.

\begin{figure}[!hbt]
\includegraphics[width=\columnwidth]{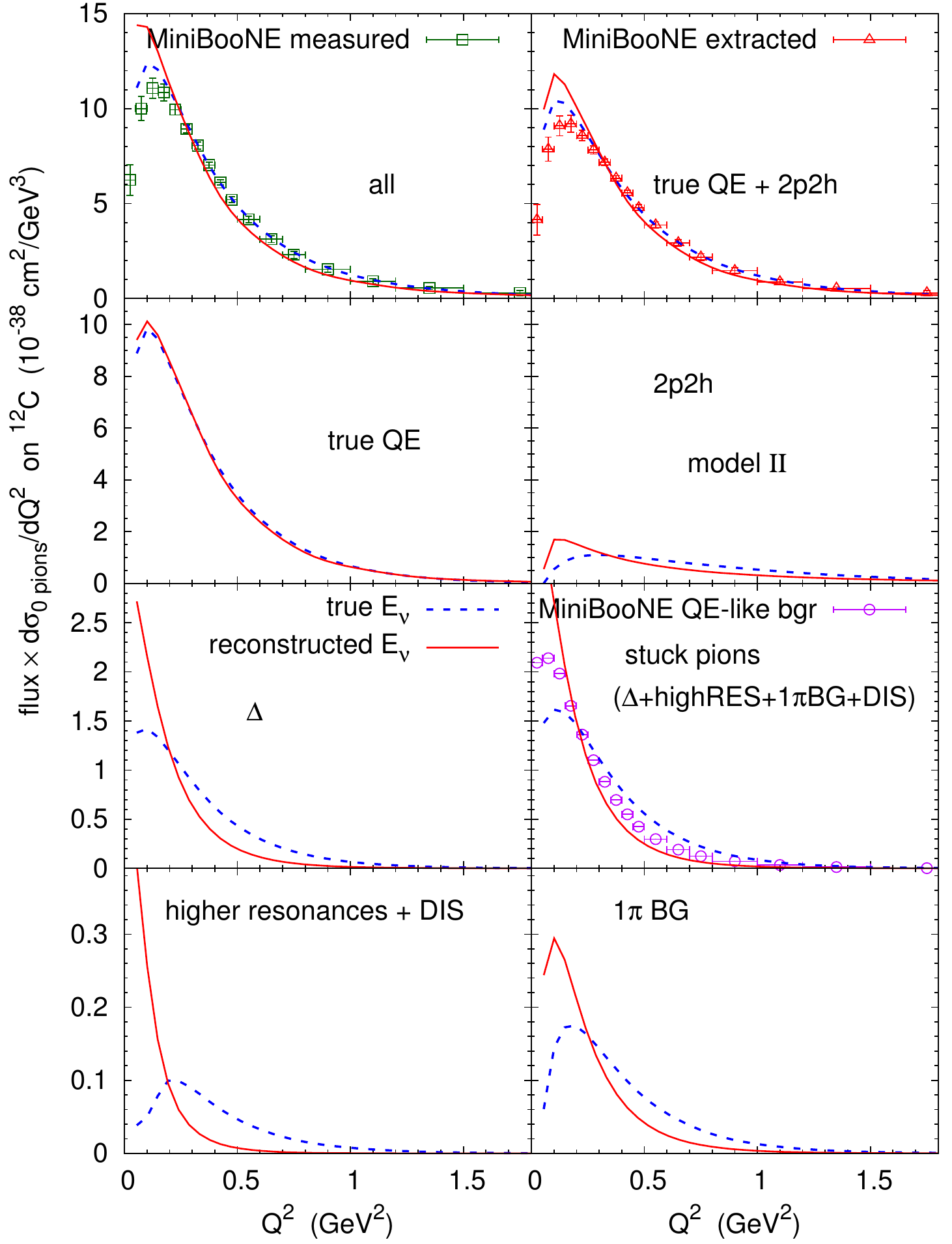}
\caption{(Color online)  The event distribution in the MiniBooNE experiment:  $\phi(E^{\rm true}) \times d\sigma_{0\pi}/dQ^2_{\rm true}$ (dashed)
vs true  and $\phi(E^{\rm rec}) \times d\tilde\sigma_{0\pi}/dQ^2_{\rm rec}$ (solid) vs reconstructed
squared momentum transfer. The data are multiplied by a factor 0.9.}
\label{fig:Q2-events-reconstruction}
\end{figure}

As we mentioned in the Introduction, the observed $Q^2$ dependence of MiniBooNE cross section can effectively be described as QE scattering with
axial mass around 1.3 GeV.
Thus, to get the MiniBooNE observed distribution, one needs a noticeable contribution which falls down  with $Q^2$
more slowly than the true-QE cross section obtained with a dipole form factor with $M_A=1\GeV$.
In our case this is a 2p2h contribution. One would naively expect, that the degree of this slowness would be quantified
by the difference between the dipoles with $M_A=1.0$ and 1.3 GeV.
The necessity of reconstructing the $Q^2$ makes this more complicated.

Figure~\ref{fig:Q2-events-reconstruction} shows the influence of the reconstruction procedure (\ref{Q2reconstr})
on the $Q^2$ distributions for the QE-like events of various origins.
Similar to the case of neutrino energy, for true-QE events distributions versus true and reconstructed energies nearly coincide.
For fake events the reconstructed distributions (solid curves) are noticeably steeper  than the true ones (dashed curves).
This leads to the same  effect for all QE-like events.
Thus, the reconstruction procedure (\ref{Q2reconstr}) makes the $Q^2$ distribution look steeper, which in turn means
that the distribution of 2p-2h contribution versus true $Q^2$ should be even flatter than the naive expectation.

Within the 2p-2h model employed in this paper, for QE and 2p-2h events (labeled ''2p2h+QE'')
the agreement of the reconstructed curve with the MiniBooNE extracted  data is not perfect.
For lower $Q^2$ the calculated curves are higher than the data; this is the region where RPA effects should bring them down \cite{Martini:2011wp, Nieves:2011pp}.
For $Q^2>0.35\GeV$ our reconstructed curve is steeper than the data.
For all events (MiniBooNE measured) the differences are larger, which is due to the different treatment
of stuck-pion  events in the GiBUU and NUANCE generators.

\section{Effects on determinations of the oscillation probability  \label{T2K} }

The T2K experiment has recently reported the first experimental observation of electron-neutrino appearance from a muon neutrino beam \cite{Abe:2011sj}
and investigated muon-neutrino disappearance \cite{Abe:2012gx} with an off-axis beam. In both studies the QE-like events were used and neutrino energy was
reconstructed assuming QE kinematics, according to Eq.~(\ref{Ereconstr}). Thus, the effects of energy reconstruction must be similar
to those in MiniBooNE experiment. In this section we show, how this influences the extraction of oscillation parameters.

\begin{figure}[!hbt]
\includegraphics[width=\columnwidth]{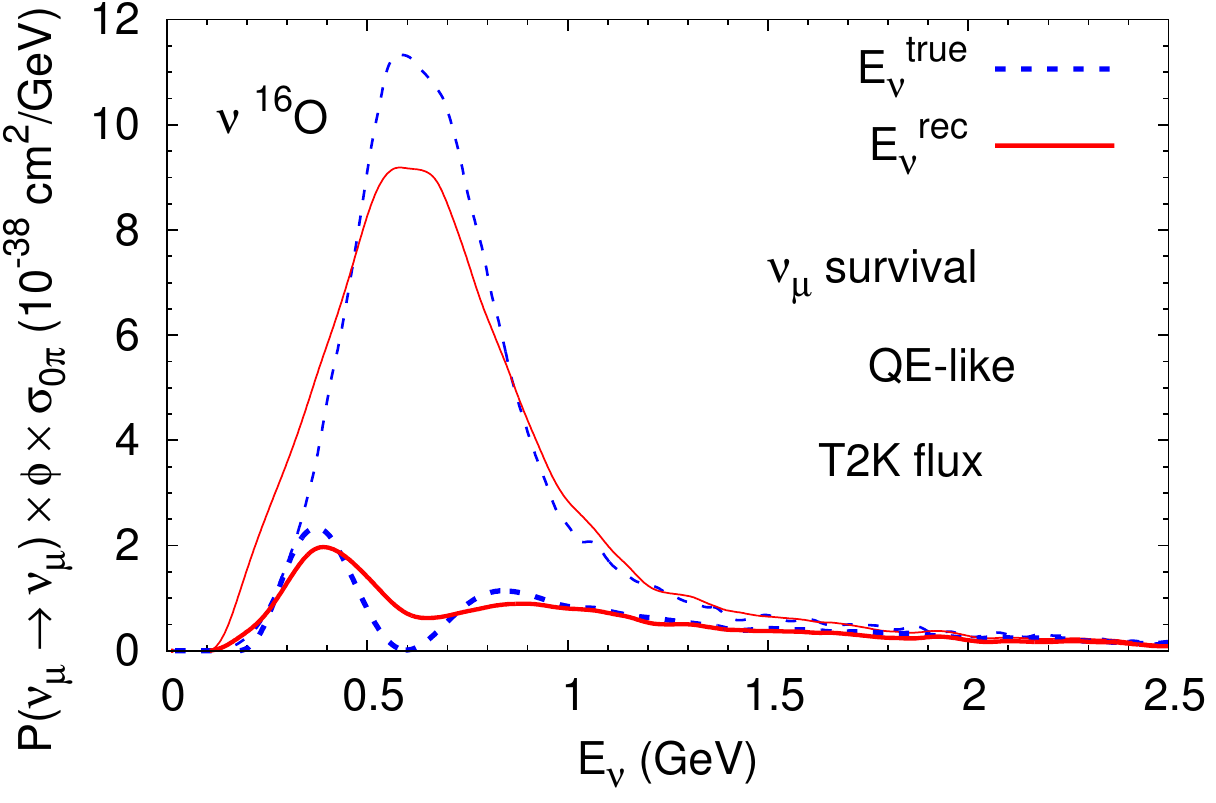}
\caption{(Color online) QE-like event distribution for the original (thin curves) and oscillated (thick curves) QE-like event distribution in T2K flux
for muon neutrino survival measurements.  The same events are shown versus true [$\phi(E^{\rm true}) \sigma_{0\pi}(E^{\rm true})$, dashed]  and reconstructed [$\phi(E^{\rm rec}) \tilde\sigma_{0\pi}(E^{\rm rec}$], solid) neutrino energy.}
\label{fig:T2Kflux-oscillations-Che-before-after-mumu}
\end{figure}
\begin{figure}[!hbt]
\includegraphics[width=\columnwidth]{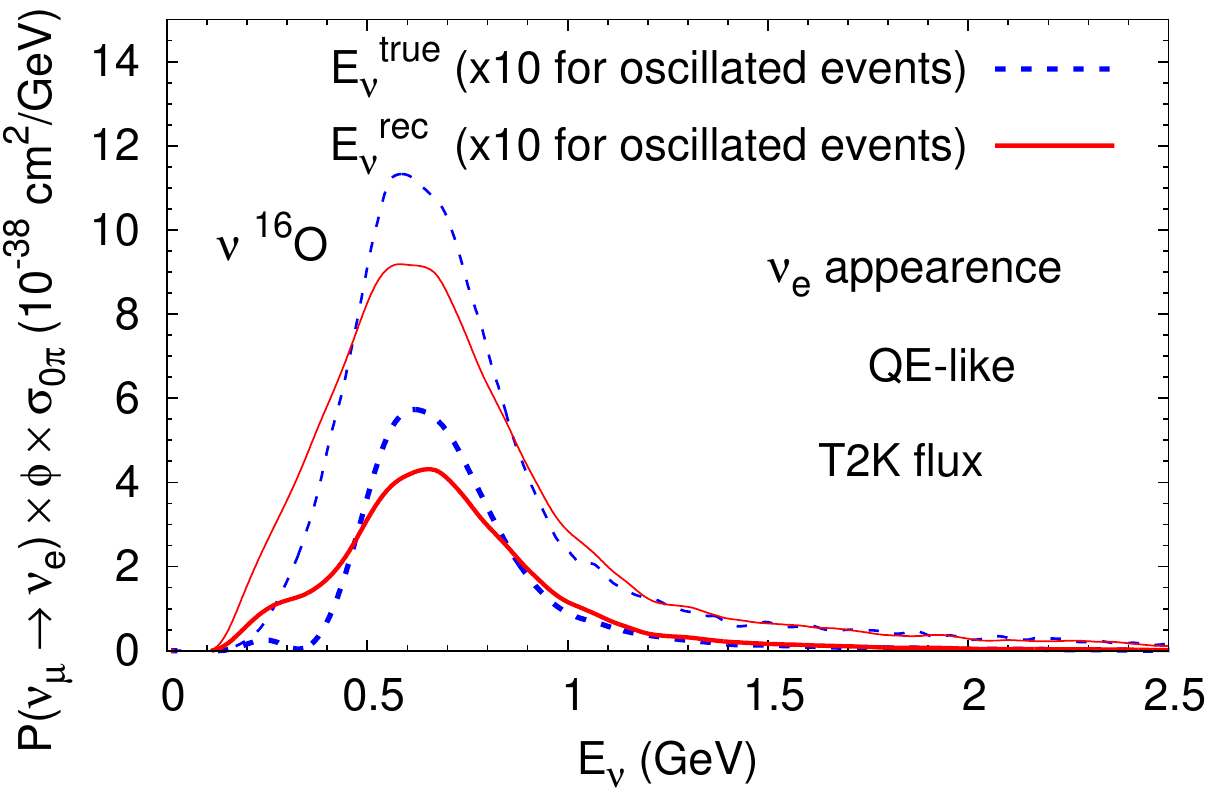}
\caption{(Color online) The same as Fig.~\ref{fig:T2Kflux-oscillations-Che-before-after-mumu}, but for electron neutrino appearance measurements. The oscillated event curves have been multiplied by a factor of 10 to enhance the visibility of their difference.}
\label{fig:T2Kflux-oscillations-Che-before-after-mue}
\end{figure}

Fig.~\ref{fig:T2Kflux-oscillations-Che-before-after-mumu} shows the actual oscillation signal plotted as a function of both the true (dashed curves) and the reconstructed (solid curves) energy for the muon-neutrino survival events.
Fig.~\ref{fig:T2Kflux-oscillations-Che-before-after-mue} show the oscillation signal for electron-neutrino appearance events under assumption $\delta_\mathrm{CP}=0$.
Thin curves show the QE-like event distribution for the original T2K flux \cite{GalymovDiss:2012}
while thick curves show that for the oscillated flux calculated with the oscillation parameters
\begin{equation}
\Delta m_{23}^2 = 2.5 \cdot 10^{-3} \eV^2 \quad \mbox{and} \quad \sin 2\theta_{23} = 1.0
\end{equation}
for the muon-neutrino survival signal,
and
\begin{equation}
\begin{array}{c}
\Delta m_{21}^2 = 7.6 \cdot 10^{-5} \eV^2,
\\
\sin 2\theta_{12} = 0.927, \quad \sin 2\theta_{13} = 0.316
\end{array}
\end{equation}
for the electron-neutrino appearance signal. These values are taken from Ref.~\cite{Bishai:2012ss} and
are very close to those extracted for $\delta_{\rm CP} = 0$ in Ref.\ \cite{Abe:2011sj}.

In the disappearance signal the oscillation minimum is significantly affected: plotted as a function of reconstructed energy the minimum is smeared out and shifted to a higher energy (by about 50 MeV). Also the first maximum is similarly affected. Correspondingly, in the electron-neutrino appearance experiment the maximum versus the true energy lies higher by about 20\%.

\begin{figure}[!hbt]
\includegraphics[width=\columnwidth]{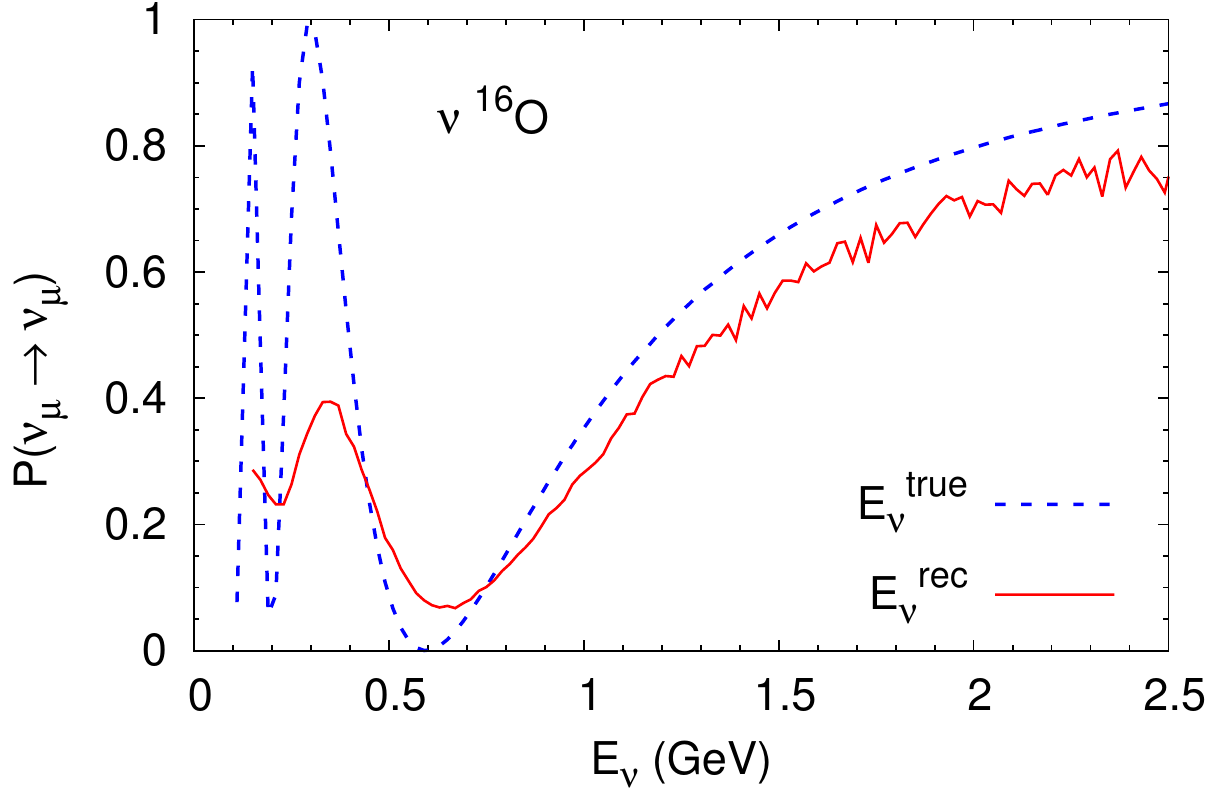}
\caption{(Color online) Muon neutrino survival probability versus true and reconstructed energy for the T2K flux.}
\label{fig:T2Kflux-oscillations-Che-ratio-mumu}
\end{figure}

\begin{figure}[!hbt]
\includegraphics[width=\columnwidth]{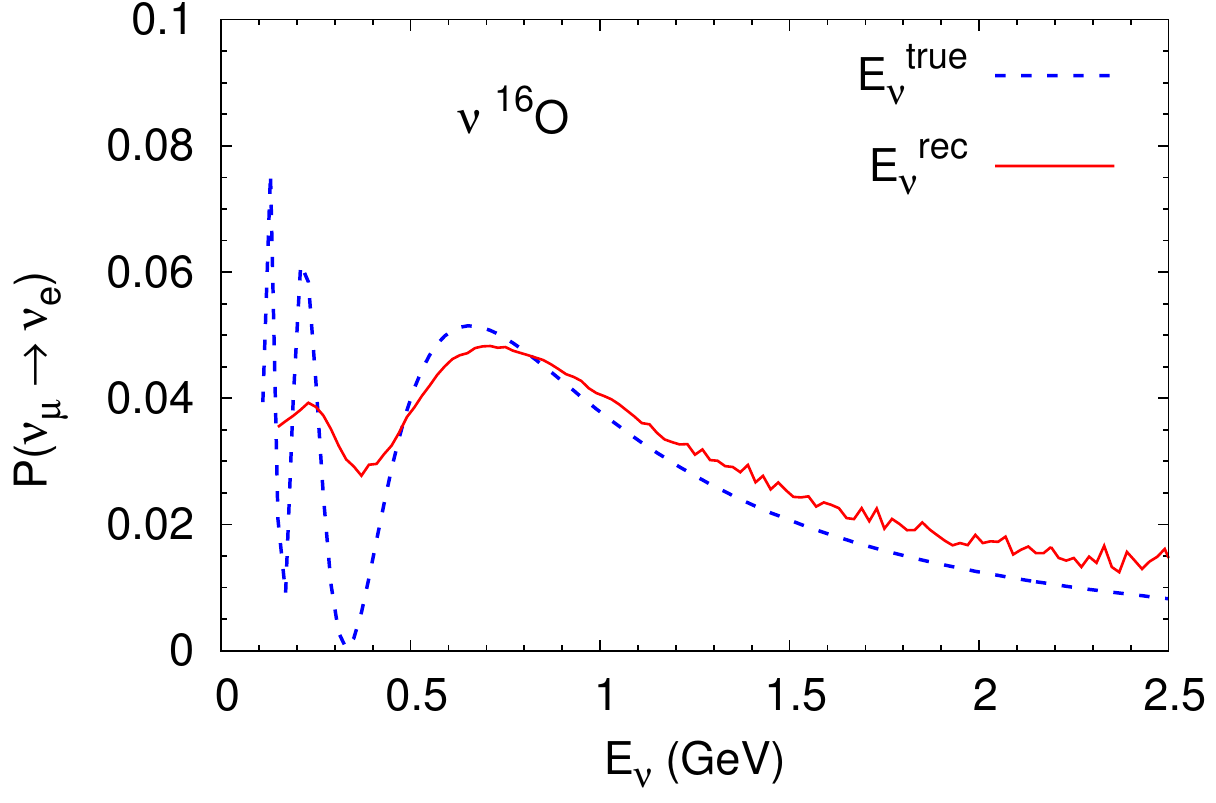}
\caption{(Color online) The same as Fig.~\ref{fig:T2Kflux-oscillations-Che-ratio-mumu}, but for electron neutrino appearance probability.}
\label{fig:T2Kflux-oscillations-Che-ratio-mue}
\end{figure}

To make these points even more visible and to demonstrate their effect on the oscillation probability we show in Figs.\ \ref{fig:T2Kflux-oscillations-Che-ratio-mumu} and \ref{fig:T2Kflux-oscillations-Che-ratio-mue} also the oscillation probability defined as the ratio of the oscillated to the initial event distributions.
Again the minimum in the curve versus reconstructed energy is smeared out and and shifted by about 50 MeV towards higher energies. The same is true for the first maximum in the electron-neutrino appearance probability.
The effect here is, however, not so large as for the event distribution. This is because the energies for the original flux
(thin solid curves in Figs.~\ref{fig:T2Kflux-oscillations-Che-before-after-mumu} and \ref{fig:T2Kflux-oscillations-Che-before-after-mue})
are shifted in a similar way as those for the oscillated flux (thick solid curves).

A quick look at the oscillation formulas shows that they depend on the combination $\Delta m^2 \times L/4E_{\nu}$, so
any uncertainty in $E_\nu$ is directly connected with a corresponding uncertainty in $\Delta m^2$. First attempts to explore also the sensitivity of the other oscillation parameters to uncertainties in the reaction mechanism have been undertaken in Ref.\ \cite{Meloni:2012fq} for the effects of 2p-2h excitations and have led to substantial effects on the mixing parameters.

\begin{figure}[!hbt]
\includegraphics[width=\columnwidth]{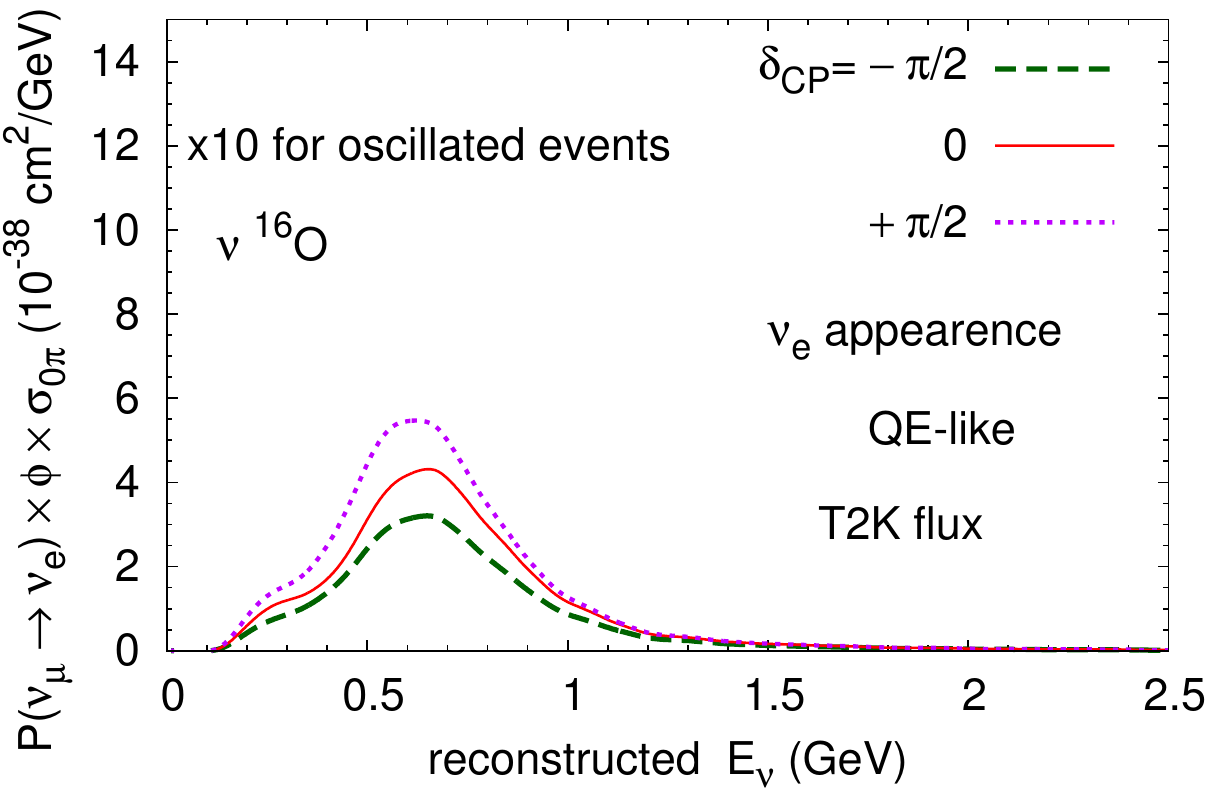}
\caption{(Color online) QE-like event distribution for the oscillated QE-like event distribution in the T2K flux
for electron neutrino appearance measurements for various CP violation phases. The oscillated event curves have been multiplied by a factor of 10 to enhance the visibility of their differences.}
\label{fig:T2Kflux-oscillations-Che-mue-deltaCP}
\end{figure}

It is interesting to compare the uncertainties in the oscillation signal that are caused by the energy reconstruction with the sensitivity of the oscillation to a $CP$-violating phase. This latter is shown in Fig.\ \ref{fig:T2Kflux-oscillations-Che-mue-deltaCP}. It is clearly seen that the sensitivity to a $CP$ violating phase is comparable to the uncertainty caused by the energy reconstruction which can be seen in Fig.\ \ref{fig:T2Kflux-oscillations-Che-before-after-mue}.

\section{Summary and Conclusions}
\label{conclusion}
The neutrino energy reconstruction using kinematics of QE scattering depends on a clean identification of true-QE events. In nuclear targets this, however, becomes difficult. Both new primary reaction mechanisms and final state interactions may cause a misidentification.

A recent example for the errors introduced by such a misidentification is provided by the energy-separated cross sections for QE-like and QE scattering obtained by the MiniBooNE experiment. Here the extracted QE scattering distributions could only be fitted by using an axial mass of $M_A = 1.3 \GeV$ and this result has lead to a discussion of possible in-medium effects and to a multitude of partly contradictory explanations. Now the general consensus is that this result was obtained because the true QE scattering mechanism was not correctly identified and, therefore, the extraction of $M_A$ was made using the wrong theoretical framework. As a consequence of this incorrect identification not only the extracted $M_A$ value was incorrect, but also the QE cross sections as a function of reconstructed energy were found to be wrong.

So far, only the QE scattering cross sections as extracted by MiniBooNE have been analyzed \cite{Martini:2012fa,Nieves:2012yz}. This extraction involved some model dependence since the so-called stuck-pion events were removed from the experimental QE-like data set by using a tuned version of the generator NUANCE. Motivated by this fact we have, in this paper, analyzed also the full QE-like cross sections in one consistent model, GiBUU, which contained all the pion production events, true QE scattering and the 2p-2h excitations. The main result is that \emph{all} other processes, not just 2p-2h, lead to a significant downward shift of the reconstructed energy. For the MiniBooNE flux this shift amounts to about 500 MeV for the 2p-2h excitations (at the peak of their event distribution), to about 500-600 MeV for pionic processes involving the $\Delta$ resonance and pion background contributions and to a shift of about 1 GeV for higher resonances and DIS. On the other hand, for true QE scattering the energy reconstruction works quite well (by construction). For the full QE-like event that involves all these reaction mechanisms the shift in energies then amounts to about 200 MeV.

Even larger is the effect on the energy-separated cross sections where the QE-like and the extracted QE cross sections have a dependence on energy that is very different for the true and the reconstructed energies. We have shown in this paper that the analysis involving all reaction mechanisms can explain both the extracted QE and the measured QE-like energy separated cross section. It also explains naturally why the difference between QE-like and QE-extracted events is quite large already at low energies and then becomes smaller toward larger neutrino energies, contrary to expectation. This unexpected behavior is again due to errors in the energy reconstruction.

We have, furthermore, shown in this paper how these uncertainties affect the analysis and the extraction of oscillation parameters from the ongoing T2K experiment. Most important here is the result that the uncertainties necessarily connected with the energy reconstruction are as large as the expected sensitivity of the oscillation result on the $CP$ invariance violating phase.

Finally, in an appendix, we have given the transformation from reconstructed to true energies in a form that may be useful for experimental analyses.

\acknowledgments
We gratefully acknowledge many inspiring discussions with Juan Nieves and Luis Alvarez-Ruso.
This work is supported by DFG, BMBF, and HIC for FAIR.

\appendix
\section{Energy reconstruction}

In an experiment only the flux averaged cross section is directly measured and by the reconstruction
method the product $\phi(E^{\rm rec}) \tilde{\sigma}_{0\pi}(E^{\rm rec})$ can be obtained.
The challenge then is to obtain the true event rate from this analysis.

The true distribution can be obtained by integrating out the reconstructed energies in Fig.\ \ref{fig:events-reconstruction3D}
\begin{equation}
\int N(E_{\nu}^\mathrm{rec},E_{\nu}^\mathrm{true})\, \dd E^{\rm rec} = \phi(E_{\nu}^\mathrm{true}) \sigma_\mathrm{0\pi}(E_{\nu}^\mathrm{true}) ~.
\label{event-distrib-true}
\end{equation}
This true event distribution can be related to the reconstructed one by
\begin{equation}
\begin{array}{l}
\displaystyle
\phi(E_{\nu}^\mathrm{true}) \sigma_\mathrm{0\pi}(E_{\nu}^\mathrm{true}) =
\\[2mm]
\displaystyle \hspace*{10mm}
=\int {\cal{P}}(E_{\nu}^\mathrm{true}|E_{\nu}^\mathrm{rec}) \phi(E_{\nu}^\mathrm{rec}) \tilde \sigma_\mathrm{0\pi}(E_{\nu}^\mathrm{rec})\, \dd E^{\rm rec} ~,
\end{array}
\end{equation}
where ${\cal{P}}(E_{\nu}^\mathrm{true}|E_{\nu}^\mathrm{rec})$ is the probability density of finding a true energy $E_{\nu}^\mathrm{true}$ in a distribution of events all having the same reconstructed energy $E_{\nu}^\mathrm{rec}$.

\begin{figure}[!hbt]
\includegraphics[width=\columnwidth]{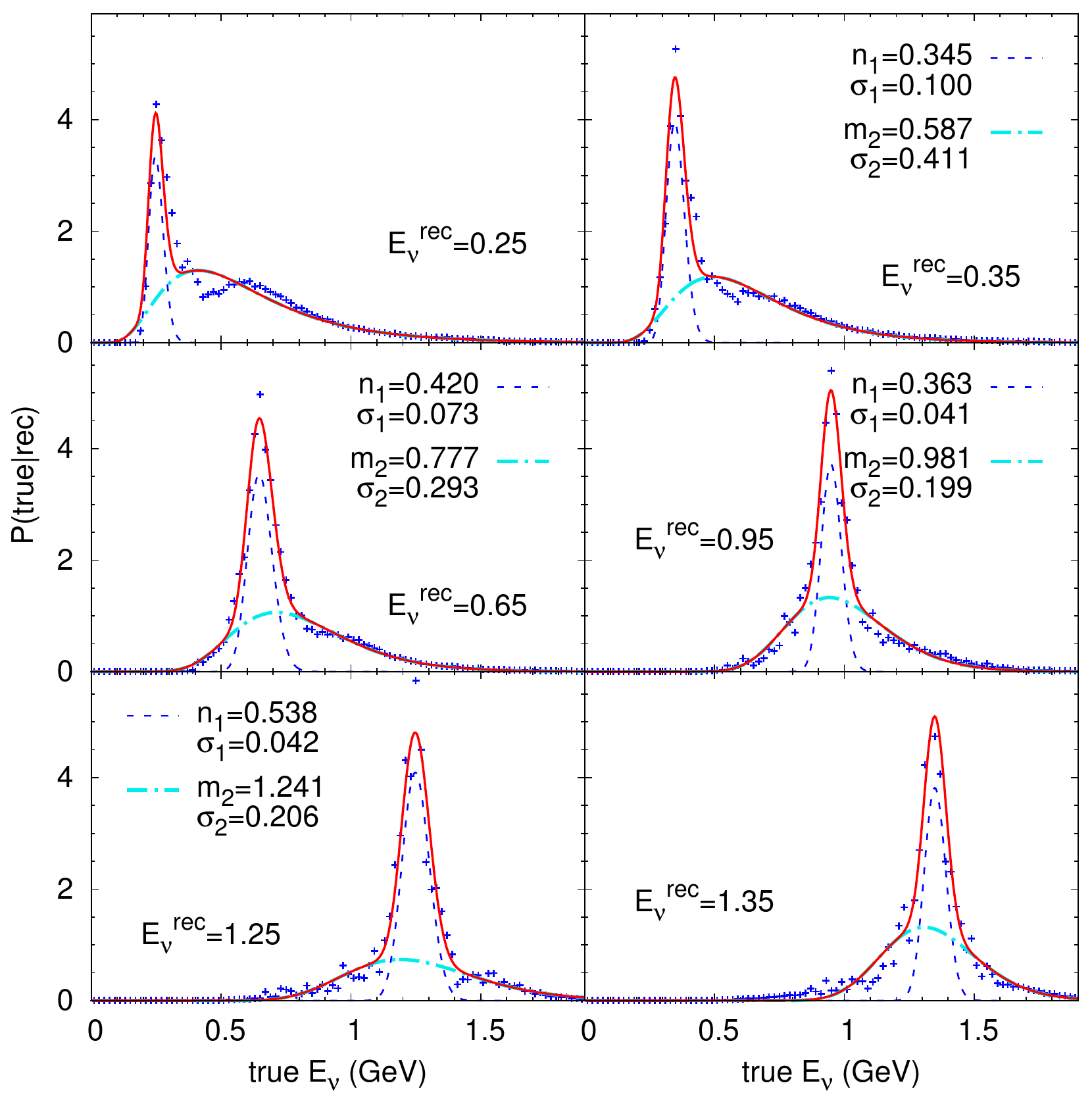}
\caption{(Color online) Conditional probability densities   ${\cal{P}}(\mathrm{true}|\mathrm{rec})$ of finding a true energy $E_\nu^\mathrm{true}$
for fixed reconstructed energies of $0.25\pm 0.01$, $0.35\pm 0.01$,  $0.65\pm 0.01$,  $0.95\pm 0.01$,  $1.05\pm 0.01$,  $1.25\pm 0.01$,  and $1.35\pm 0.01 \GeV$.
The fits are according to Eq.~(\ref{loggauss-loggauss}). }
\label{fig:prob-true-rec}
\end{figure}

From this equation ${\cal{P}}(\mbox{true}|\mbox{rec})$ can be read off. It is given by
\begin{equation}
\begin{array}{lll}
 \displaystyle
 {\cal{P}}(\mbox{true}|\mbox{rec}) &=& {\cal{P}}(E_{\nu}^\mathrm{true}|E_{\nu}^\mathrm{rec})
\\[2mm]
  &=& \frac1{\phi(E_{\nu}^\mathrm{rec})  \tilde{\sigma}_\mathrm{0\pi}(E_{\nu}^\mathrm{rec})}\,
  N(E_{\nu}^\mathrm{rec},E_{\nu}^\mathrm{true}) \ .
\end{array}
\label{Prob-true-rec}
\end{equation}
Here $\phi(E^{\rm rec})$ stands for the true flux, read off at the reconstructed energy, and the QE-like cross section versus reconstructed energy,  $\tilde{\sigma}_\mathrm{0\pi}(E_{\nu}^\mathrm{rec})$, is defined by Eq.\ (\ref{event-distrib-rec}) and shown in Fig.~\ref{fig:sigma-reconstruction}.
In Eq.~(\ref{Prob-true-rec}) $E_{\nu}^\mathrm{rec}$  is a parameter and $E_{\nu}^\mathrm{true}$ a variable, so that
the probability density is normalized as
\[
 \int {\cal{P}}(\mbox{true}|\mbox{rec})\, \dd E_{\nu}^\mathrm{true} = 1 ~ .
\]

\begin{table}[!htb]
\caption{Fit of  ${\cal{P}}(\mbox{true}|\mbox{rec})$ according to Eq.~(\ref{loggauss-loggauss}). The parameters are appropriate for the MiniBooNE flux and
a C target.}
\[
\begin{array}{lcccc}
\hline
E_\nu (\mathrm{GeV})	&	n_1	& \sigma_1 (\mathrm{GeV}) & m_2 (\mathrm{GeV}) & \sigma_2 (\mathrm{GeV})
\\
\hline
\\
0.35 \qquad &	0.345	& 0.100			  & 0.587	       & 0.411
\\
0.45	&	0.365	& 0.085			  & 0.637	       & 0.361
\\
0.55	&	0.417	& 0.078			  & 0.719	       & 0.321
\\
0.65	&	0.420	& 0.073			  & 0.777	       & 0.293
\\
0.75	&	0.397	& 0.059			  & 0.844	       & 0.259
\\
0.85	&	0.323	& 0.038			  & 0.894	       & 0.199
\\
0.95	&	0.363	& 0.041			  & 0.981	       & 0.199
\\
1.05	&	0.427	& 0.043			  & 1.083	       & 0.185
\\
1.15	&	0.496	& 0.046			  & 1.141	       & 0.189
\\
1.25	&	0.538	& 0.042			  & 1.241	       & 0.206
\\
\hline
  \end{array}
\]
\label{tab:prob-true-rec}
\end{table}

The distribution ${\cal{P}}(\mathrm{true}|\mathrm{rec})$ is of primary interest for neutrino experiments. For fixed values of true and reconstructed energy,
its connection with the
${\cal{P}}(\mathrm{rec}|\mathrm{true})$ can be obtained by comparing Eqs.\ (\ref{Prob-rec-true}) and (\ref{Prob-true-rec}),
\begin{equation}
{\cal{P}}(\mathrm{true}|\mathrm{rec})= \frac{\phi(E_{\nu}^\mathrm{true})\,\sigma_\mathrm{0\pi}(E_{\nu}^\mathrm{true}) }{\phi(E_{\nu}^\mathrm{rec})\,\tilde{\sigma}_\mathrm{0\pi}(E_{\nu}^\mathrm{rec})}\, {\cal{P}}(\mbox{rec}|\mbox{true}) ~.
\end{equation}

Contrary to the ${\cal{P}}(\mathrm{rec}|\mathrm{true})$ discussed in Sec.~\ref{energy-QEKinematics}, ${\cal{P}}(\mathrm{true}|\mathrm{rec})$ depends on the specific neutrino flux and is thus different for different experiments even if they use identical techniques for event identification.
The probability densities (\ref{Prob-true-rec}) calculated using the MiniBooNE flux are shown in Fig.~\ref{fig:prob-true-rec}  for various reconstructed energies.  We have found that in the energy region $0.35 <E_\nu^\mathrm{rec}<1.25 \GeV$ they can be described quite well by a sum of two lognormal distributions
\begin{equation}
\begin{array}{lll}
\displaystyle
{\cal{P}}(\mbox{true}|\mbox{rec}) &=& \displaystyle \frac1{\sqrt{2\pi}}\left[
  \frac{n_1}{x \sigma_1} \exp\left ( - \frac{(\ln(x)-\ln(m_1))^2}{2 \sigma_1^2} \right) \right. \hspace*{1mm}
\\[5mm]
&+& \left. \displaystyle \frac{1-n_1}{x \sigma_2} \exp\left ( - \frac{(\ln(x)-\ln(m_2))^2}{2 \sigma_2^2} \right)
\right] ~.
\end{array}
\label{loggauss-loggauss}
\end{equation}
Here $x=E_\nu^\mathrm{true}$ is a variable, $m_1=E_\nu^\mathrm{rec}$ is the center of the first peak, which (as discussed above) mainly
comes form Fermi motion in true-QE scattering. The second much broader peak is due to the 2p-2h and stuck-pion QE-like
processes. At low $E_\nu^\mathrm{rec}$ it is centered at noticeably higher energies then the first one;
with increasing $E_\nu^\mathrm{rec}$ it shifts closer and closer to the first peak and eventually to the left of the first peak.
Parameters $n_1$, $\sigma_1$, $m_2$, and $\sigma_2$ are fitted.
In Fig.~\ref{fig:prob-true-rec} the best fit is shown by the solid line, the first term in Eq.~(\ref{loggauss-loggauss}) by the dashed line
and the second one by the dash-dotted line.

For other energies the fit parameters are given in Table~\ref{tab:prob-true-rec}.

For lower, $E_\nu^\mathrm{rec}<0.25 \GeV$,  and higher reconstructed energies, $E_\nu^\mathrm{rec}>1.35 \GeV$,
the shape of the probability distribution differs from the functional form  (\ref{loggauss-loggauss}). This is also
illustrated in Fig.~\ref{fig:prob-true-rec}.

\bibliographystyle{apsrev}
\bibliography{nuclear}

\end{document}